\documentclass[journal]{IEEEtran}

\usepackage{amsmath}
\usepackage{amsfonts,amssymb}
\usepackage{graphicx}
\allowdisplaybreaks[4]
\usepackage{xcolor}
\definecolor{deepred}{RGB}{205,38,38}
\usepackage{algorithm}
\usepackage{algorithmic}
\usepackage{amsthm}

\usepackage{lineno,hyperref}
\modulolinenumbers[5]
\usepackage{bm}

\usepackage{multirow}
\usepackage{array}
\usepackage{booktabs}
\usepackage{lipsum}
\usepackage{xcolor}
\usepackage{etoolbox}
\ifCLASSINFOpdf
\else
\fi
\ifCLASSOPTIONcompsoc
  \usepackage[caption=false,font=normalsize,labelfont=sf,textfont=sf]{subfig}
\else
  \usepackage[caption=false,font=footnotesize]{subfig}
\fi

\graphicspath{{./FiguresJournal/}} 

\begin{document}

\title{Input Convex Neural Network-Assisted  Optimal Power Flow in Distribution Networks: Modeling, Algorithm Design, and Applications} 
\author{Rui Cheng,~\IEEEmembership{Member,~IEEE}, Yuze Yang,~\IEEEmembership{Student Member,~IEEE},
Wenxia Liu,~\IEEEmembership{Member,~IEEE},
Nian Liu,~\IEEEmembership{Member,~IEEE}, Zhaoyu Wang,~\IEEEmembership{Senior Member,~IEEE}
}

\maketitle
\begin{abstract} 
This paper proposes an  input convex neural network (ICNN)-Assisted optimal power flow (OPF) in distribution networks. Instead of relying purely on optimization or machine learning,  the ICNN-Assisted OPF is a combination of optimization and machine learning. It utilizes ICNN to learn the nonlinear but convex mapping from control variables to system state variables, followed by embedding into constrained optimization problems as convex constraints. Utilizing a designed ICNN structure, a fast primal-dual gradient method is developed to solve the ICNN-Assisted OPF, with the chain rule of deep learning applied to accelerate the algorithmic implementation. Convergence and optimally properties of the algorithm design are further established. Finally, different distribution network applications are discussed and proposed by means of the ICNN-Assisted OPF.
\end{abstract}


\section{Introduction}
\label{sec:Intro}
\IEEEPARstart{O}ptimal power flow problems play a key role in distribution network decision-making, operation, and planning, which have attracted  massive research efforts in recent decades. Traditionally, distribution network OPF relies heavily on accurate distribution network models, and then resorts to different optimization algorithms and methods, e.g., Newton method, interior point method \cite{CMB}-\cite{FC}. However, accurate distribution network modeling is  difficult and challenging.

Fortunately, distribution networks are undergoing a fundamental architecture transformation to become more intelligent, controllable, and open, thanks to  the increasing deployment of communication, computing, information devices, such as advanced metering infrastructure and smart meter \cite{PNM}. Such a transformation provides large volumes of distribution network operation data, it has gained increasing attention taking advantage of data-driven and machine learning methods \cite{BH} to facilitate the implementation of distribution network OPF.

To this end, this study proposes a novel Input Convex Neural Network (ICNN)-Assisted OPF strategy by making full use of data to reduce dependence on distribution network modeling. As discussed more carefully in subsequent sections, this proposed ICNN-Assisted OPF  has three important advantages compared to many previously distribution network OPF works.

First, the general form of the proposed ICNN-Assisted OPF is applicable for distribution networks that are either \textit{unbalanced or balanced}, and in either \textit{meshed or radial} form in \textit{an unified manner}, instead of \textit{structure-dependent manner}. And it can also be applied to \textit{different research areas and topics}, e.g., economic dispatch and voltage control, by considering different objective functions and constraints.  

Second, the ICNN-Assisted OPF leverages ICNN \cite{ICNN}, a type of convex neural network, to learn the mapping from control variables to {network states}, thus constructing convex system-wide constraints. These constraints, represented by ICNN, are embedded into the convex optimization formulation, making the ICNN-Assisted OPF \textit{a combination of optimization and machine learning}. It reduces dependence on distribution network modeling while enhancing the credibility  of data-driven decision-making.

Third, a subtle structure of neural network, including the weights and Softplus activation functions, is designed for ICNN to facilitate our algorithm design. A fast primal-dual gradient algorithm \cite{JKA} is further developed to solve the ICNN-Assisted OPF in \textit{a tractable and trustworthy way} with \textit{performance guarantees}, where \textit{the chain rule of deep learning} is utilized to speed up the algorithmic implementation.  \textit{Convergence and optimality properties} of algorithm design are  established and provided in detail. 

Remaining sections are organized as follows. The relationship of this study to previous distribution network OPF studies is discussed in Section \ref{sec:Relation}. Distribution network OPF problem and its general modeling of ICNN-Assisted OPF are  described in Sections \ref{sec:OPF}-\ref{section:ICNN}.  The algorithm design of ICNN-Assisted OPF, associated with its convergence and optimality properties, is established in Section \ref{sec:algorithm}. Section \ref{sec:application} discusses different network applications by means of ICNN-Assisted OPF. And Section \ref{sec:case} reports numerical test cases that demonstrate these properties in more concrete form. The concluding Section \ref{sec:conclusion} discusses ongoing and planned future studies.

\section{Relationship to Existing Literature}
\label{sec:Relation}
Distribution network OPF has been widely investigated in the academia and industry. Conventional mathematical methods, e.g., interior point method, genetic algorithms and particle swarm
optimization \cite{SG}-\cite{NM}, are first proposed to directly solve non-convex and non-linear distribution network OPF. However, solving non-convex and non-linear optimization problems is always time-consuming and NP-hard to find global/local optimal solutions.  

To address the challenges, convex relaxation and linearization approaches have been greatly studied in the literature. Different types of convex relaxation approaches are applied to solve distribution network OPF, including but not limited to second-order cone program relaxations \cite{RAJ}-\cite{BFM}, semidefinite relaxations \cite{XB}-\cite{EDAHZ}, chordal relaxations \cite{MAS}. Compared to convex relaxation, linearization approaches \cite{MB}-\cite{RCZ} approximate non-linear power flow equations based on some assumptions, thus exhibiting lower complexity and higher computational efficiency. Nevertheless, all these works are essentially model-based, relying on accurate distribution network modeling.

Recently, researchers and practitioners are exploring a variety of data-driven and machine learning methods to solve distribution network OPF due to rich data in modern distribution networks. They make full use of data to learn the mapping from network states and environment to distribution network OPF solutions by means of supervised learning \cite{DeepAL}-\cite{DDOPF}, unsupervised learning \cite{UL} and reinforcement learning \cite{DRL}-\cite{QZK}, thus greatly improving the computational speed and efficiency.

Despite the promising results achieved by convex relaxation, linearization, and machine learning approaches, these previously proposed distribution network OPF works leave three open critical issues. First, most works lack the generalization capability, limited to specific distribution network OPF setting. More specifically, convex relaxation performance highly depends on \textit{specific distribution network structures} and \textit{objective functions}. There is no general convex relaxation for distribution network OPF. Linearization is easy to implement, but it fails to capture the nonlinear power flow properties and still highly relies on distribution network modeling.

Second, most distribution network OPF problems are purely either \textit{optimization-based} or \textit{machine learning-based}. Optimization-based distribution network OPF highly relies on distribution network model accuracy, not easy to obtain in the real world. In contrast, machine learning-based distribution network OPF does not depend on distribution network modeling, but the resulting solutions are prone to in-feasibility and cannot guarantee the satisfaction of network constraints. Furthermore, distribution system operators always lack confidence in machine learning-based distribution network OPF due to its black-box nature. 


Third, most of these works are faced with \textit{scalability, convergence, and optimality challenges}. For example, Refs.\cite{QZK2}-\cite{WW} utilize safe reinforcement learning to ensure the reliable operation of distribution networks, but these works do not provide theoretical analyzes and proofs in terms of convergence and optimality. Refs.\cite{AVG}-\cite{IM} train  binary-classification neural networks to judge the feasibility of OPF solutions, where the MLP is reformulated into mixed-integer constraints to replace complex constraints. Ref.\cite{TCL} leverages neural networks to express the degree of power flow constraint violations, then a linearization technique is further developed to convert them into  mixed-integer constraints, leading to a mixed-inter linear programming (MILP) problem. However, mixed-integer constraints, generated by neural networks, introduce numerous integer variables into optimization problems due to high-dimension and deep-layer properties of neural networks, posing \textit{the curse of dimensionality  and scalability problems}.

As carefully established in subsequent sections, the ICNN-Assisted OPF for distribution networks proposed in the current study addresses all three of these critical issues. The proposed ICNN-Assisted OPF is \textit{convex} and \textit{applicable} for different distribution network structures in \textit{an unified manner}. It is \textit{a combination of optimization and machine learning}, which is {model-free} but also {resorts to constrained optimization algorithms to increase the solution credibility}. Finally, a fast primal-dual gradient algorithm, \textit{accelerated by the chain rule of deep learning}, is proposed for the ICNN-Assisted OPF to \textit{handle scalability, convergence, and optimality challenges with theoretical guarantees}.

The previous works closest to the current study  are  \cite{ICNNVolt}-\cite{DRVVC}. All these works utilize ICNN to learn the mapping from the control variables to voltage deviations in single-phase radial distribution networks, then replace the objective function of voltage deviations with the nonlinear convex function of ICNN to perform optimization. However, these works treat the voltage constraint of radial single-phase distribution networks as a soft penalty, which do not meet the hard voltage and power flow constraints. They are applicable to voltage regulation problems without providing theoretical analyses of convergence and optimality. Instead, our proposed ICNN-Assisted OPF considers the voltage constraints and line congestion constraints as hard constraints in more complex distribution network structures, e.g., unbalanced three-phase distribution networks, and is suitable for different distribution network applications. Theoretical analyses of convergence and optimality are also established and provided.

\section{Distribution Network OPF Problem Formulation}
\label{sec:OPF}
Without loss of generalization, distribution network OPF problems can be expressed as follows:
\begin{subequations}\label{eq:OriginalOPF}
\begin{align}
    \min~~&h(\bm{p}^c,\bm{q}^c)\\
    \text{subject to:}~~&
    \bm{V}=f(\bm{p},\bm{q})\\
    &\bm{P}=g(\bm{p},\bm{q})\\
    & \bm{{V}}_{min}\leq\bm{V}\leq\bm{{V}}_{max}\\
    & \bm{{P}}_{min}\leq\bm{P}\leq\bm{{P}}_{max}\\
    & \bm{p}=\bm{p}^c+\bm{p}^u\\
    & \bm{q}=\bm{q}^c+\bm{q}^u\\
&\bm{p}^c\in\mathcal{P}^c\\
&\bm{{q}}^c\in\mathcal{Q}^c
\end{align}
\end{subequations}
where functions $\bm{V}=f(\bm{p},\bm{q})$ and $\bm{P}=g(\bm{p},\bm{q})$ represent the nonlinear power flow for \textit{radial/meshed balanced (single-phase)/unbalanced (multi-phase)} distribution networks, $\bm{p}$ and $\bm{q}$ are net real and reactive power vectors of bus,  $\bm{p}^c$ and $\bm{q}^c$ are controllable real and reactive power vectors of bus,  $\bm{p}^u$ and $\bm{q}^u$ are uncontrollable real and reactive power vectors of bus, $\bm{V}$ are the bus voltage vector.  There are two main limitations on $\bm{V}=f(\bm{p},\bm{q})$ and $\bm{P}=g(\bm{p},\bm{q})$: (1) they are nonlinear, leading to non-convex optimization; (2) they usually require accurate and detailed distribution network information, e.g., the distribution network topology and the line segment resistance/reactance, which is difficult to accurately access/obtain.

To overcome the first limitation, convex relaxation and linearization approaches for the nonlinear power flow are investigated. However, convex relaxation and linearization approaches still require accurate and detailed power network information. In addition,  convex relaxation and linearization approaches highly rely on the structure of distribution networks.  Different structures of distribution networks,  e.g., single-phase or multi-phase networks, radial or meshed networks, always necessitate different convex relaxation and linearization approaches. To handle the second limitation,  there are various data-driven model-free solutions to solve  OPF problems. However, those data-driven solutions are usually black-box solutions that cannot give the distribution network operator the confidence to utilize them in the real world.

To this end,  Section.\ref{section:ICNN} describes an ICNN-Assisted OPF in distribution networks, which focuses on embedding the ICNN into classical OPF problems by  combining conventional optimization with machine learning.

\section{ICNN-Assisted OPF Modeling in Distribution Networks}
\label{section:ICNN}
\subsection{Input Convex Neural Network: Overview}
\begin{figure}
    \centering
    \includegraphics[width=3.5in]{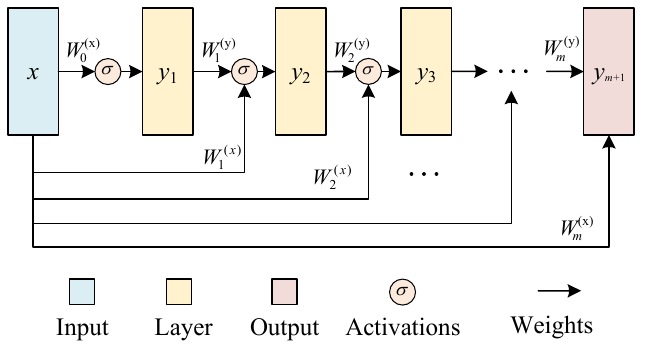}
    \caption{ICNN structure illustration}
    \label{fig:ICNN}
\end{figure}

ICNN is a neural network whose outputs are convex with respect to inputs by subtlety designing the weights and activation functions of neural network. Consider a $\{m+1\}-$layer  fully connected neural network. As shown in Fig.\ref{fig:ICNN}, this model defines a neural network over input $\bm{x}$ using the following architecture:
\begin{subequations}
\begin{align} \bm{y}_{i+1}&=\delta_i\big(\bm{W}_i^{(y)}\bm{y}_i+\bm{W}_i^{(x)}\bm{x}+\bm{b}_i\big),
\forall i=0,...,m-1\\
\bm{y}_{m+1}&=\bm{W}_m^{(y)}\bm{y}_m+\bm{W}_m^{(x)}\bm{x}+\bm{b}_m=f(\bm{x};\bm{\theta})
\end{align}
\end{subequations}
where $\delta_i$ denotes the layer activation (with $\bm{y}_0, \bm{W}_0^{(y)}=\bm{0}$),  $\bm{\theta}=\{\bm{W}_{0:m}^{(y)},\bm{W}_{0:m}^{(x)},\bm{b}_{0:m}\}$ are the parameters, and $\delta_i$ is a non-linear activation function of $i$-th layer. 

It is proved in \cite{ICNN} that the function $f$ is convex in $\bm{x}$ provided that all $\bm{W}_{1:m}^{(y)}$ are non-negative, and all activation functions $\delta_i$ for each layer are convex and non-decreasing. The proof is simple and follows from the fact that non-negative sums of convex functions are also convex and that the composition of a convex and convex non-decreasing function is also convex.
\subsection{Constraint Replacement by ICNN}
One intuitive idea is to learn the mapping from the input $\bm{p},\bm{q}$ to the output $\bm{V}$ and $\bm{P}$ by means of ICNN, and then replace (\ref{eq:OriginalOPF}b)-(\ref{eq:OriginalOPF}c)
 , i.e., $\bm{V}=h(\bm{p},\bm{q})$ and $\bm{P}=g(\bm{p},\bm{q})$, by ICNN functions $\bm{V}=f(\bm{p},\bm{q};\bm{\tilde{\theta}}_V)$ and $\bm{P}=g(\bm{p},\bm{q};\bm{\tilde{\theta}}_P)$ in the OPF problem. However, such a replacement still leads to a non-convex OPF problem due to the nonlinear properties of equations $\bm{V}=f(\bm{p},\bm{q};\bm{\tilde{\theta}}_V)$ and $\bm{P}=g(\bm{p},\bm{q};\bm{\tilde{\theta}}_P)$
even though $\bm{V}=f(\bm{p},\bm{q};\bm{\tilde{\theta}}_V)$ and $\bm{P}=g(\bm{p},\bm{q};\bm{\tilde{\theta}}_P)$ are convex functions.

Note that (\ref{eq:OriginalOPF}d)-(\ref{eq:OriginalOPF}e) can be rewritten as follows:
\begin{subequations}
\begin{align}
    \Big|\bm{V}-\frac{\bm{{V}}_{min}+ \bm{{V}}_{max}}{2}\Big|\leq\Delta{\bm{V}}=\frac{\bm{{V}}_{max}-\bm{{V}}_{min}}{2}\\
    \Big|\bm{P}-\frac{\bm{{P}}_{min}+ \bm{{P}}_{max}}{2}\Big|\leq\Delta{\bm{P}}=\frac{\bm{{P}}_{max}-\bm{{P}}_{min}}{2}
\end{align}
\end{subequations}
Instead of learning the mapping from the input $\bm{p},\bm{q}$ to the output $\bm{V}$ and $\bm{P}$, we utilize ICNN to learn the nonlinear and convex functions $f(\bm{p},\bm{q};\bm{\tilde{\theta}}_V)$
and $g(\bm{p},\bm{q};\bm{\tilde{\theta}}_P)$, reflecting the mapping from the input $\bm{p},\bm{q}$ to $\Big|\bm{V}-\frac{\bm{{V}}_{min}+ \bm{{V}}_{max}}{2}\Big|$ and $\Big|\bm{P}-\frac{\bm{{P}}_{min}+ \bm{{P}}_{max}}{2}\Big|$. That is:
\begin{subequations}
\begin{align}
    f(\bm{p},\bm{q};\bm{\tilde{\theta}}_V)\approx\Big|\bm{V}-\frac{\bm{{V}}_{min}+ \bm{{V}}_{max}}{2}\Big|\\
    g(\bm{p},\bm{q};\bm{\tilde{\theta}}_P)\approx\Big|\bm{P}-\frac{\bm{{P}}_{min}+ \bm{{P}}_{max}}{2}\Big|
\end{align}
\end{subequations}
where $\tilde{\theta}_V$ and $\tilde{\theta}_P$ are the ICNN parameters. After this replacement, the ICNN-Assisted OPF can be expressed as follows: 

\noindent
\textbf{ICNN-Assisted OPF:}
\begin{subequations}\label{eq:ICNNOPF}
\begin{align}
    \min~~&h(\bm{p}^c,\bm{q}^c)\\
    \text{subject to:}~~
    & f(\bm{p},\bm{q};\bm{\tilde{\theta}}_V)\leq\Delta{\bm{V}}\\
    &g(\bm{p},\bm{q};\bm{\tilde{\theta}}_P)\leq\Delta{\bm{P}}\\
    & \bm{p}=\bm{p}^c+\bm{p}^u\\
    & \bm{q}=\bm{q}^c+\bm{q}^u\\
&\bm{{p}}^c\in\mathcal{P}^c\\
&\bm{{q}}^c\in\mathcal{Q}^c
\end{align}
\end{subequations}

\medskip
\noindent
[\textbf{Proposition 1}] Suppose the following conditions hold for the \textbf{ICNN-Assisted OPF}:
\begin{itemize}
    \item \textbf{[P1.A]} $h(\bm{p}^c, \bm{q}^c)$ is convex;
    \item \textbf{[P1.B]} $f(\bm{p},\bm{q};\bm{\tilde{\theta}}_V)$ and 
    $g(\bm{p},\bm{q};\bm{\tilde{\theta}}_P)$ are
    convex with respect to $\bm{p},\bm{q}$.
    \item \textbf{[P1.C]} $\mathcal{P}^c$ and $\mathcal{Q}^c$ are convex sets. 
\end{itemize}
Then {ICNN-Assisted OPF} is a convex optimization problem. 

\medskip
\noindent
[\textbf{Proof of Proposition 1}] From \textbf{[P1.A]}, it following the objective function of the \textbf{ICNN-Assisted OPF} is convex. A function is convex if and only if its epigraph is a convex set. Thus, from \textbf{[P1.B]}, we know that $f(\bm{p},\bm{q};\bm{\tilde{\theta}}_V)$ and $g(\bm{p},\bm{q};\bm{\tilde{\theta}}_P)$ are convex with respect to $\bm{p},\bm{q}$ if and only if (\ref{eq:ICNNOPF}b) and  (\ref{eq:ICNNOPF}c) are convex sets. [\textbf{P1.C}] shows (\ref{eq:ICNNOPF}f) and (\ref{eq:ICNNOPF}g) are convex sets. In addition, the equality constraint functions (\ref{eq:ICNNOPF}d) and (\ref{eq:ICNNOPF}e) are affine. Thus,  we can say the \textbf{ICNN-Assisted OPF} is a convex optimization problem.

The \textbf{ICNN-Assisted OPF} exhibits the following characteristics:
\begin{itemize}
    \item \textbf{The ICNN-Assisted OPF is convex.} Due to the convex characteristics of ICNN, $f(\bm{p},\bm{q};\bm{\tilde{\theta}}_V)$ and $g(\bm{p},\bm{q};\bm{\tilde{\theta}}_P)$ are all convex functions with respect to $\bm{p}$, $\bm{q}$. Thus, the ICNN-Assisted OPF includes the convex objective function and convex constraint set, making itself a convex program. 
    \item  \textbf{The ICNN-Assisted OPF is a combination of optimization and machine learning.} With respect to $f(\bm{p},\bm{q};\bm{\tilde{\theta}}_V)$ and $g(\bm{p},\bm{q};\bm{\tilde{\theta}}_P)$, they do not require the distribution network information. However, rather than relying on purely data-driven model-free solutions, the ICNN-Assisted OPF combines optimization and machine learning with consideration voltage and line congestion limits.
    \item \textbf{The ICNN-Assisted OPF is generalized.} The ICNN-Assisted OPF can be applied to different research areas and topics, e.g., economic dispatch and voltage control, by choosing different $h(\bm{p}^c,\bm{q}^c)$. The ICNN-Assisted OPF can also be applied to different network structures, e.g., single-phase or three-phase networks, radial or meshed networks,
    by providing the corresponding dataset. 
\end{itemize}

\section{ICNN-Assisted OPF: Algorithm Design}
\label{sec:algorithm}
\subsection{Activation Function Selection}
It should be pointed out that solving the ICNN-Assisted OPF (\ref{eq:ICNNOPF}) could be extremely challenging even though it is a convex program, where $f(\bm{p},\bm{q};\bm{\tilde{\theta}}_V)$ and $g(\bm{p},\bm{q};\bm{\tilde{\theta}}_P)$  could be high-dimension, non-linear, and non-smooth due to numerous neurons and deep layers in the ICNN. In this case, it is of great importance to select appropriate activation functions to facilitate the algorithm design.

\medskip
\noindent
[\textbf{Remark 1}] One popular convex and non-decreasing activation function is a ReLU function. However, the ReLu function is piecewise linear and non-smooth, which always introduces integer variables to express its analytical formulation. It indicates$f(\bm{p},\bm{q};\bm{\tilde{\theta}}_V)$ and $g(\bm{p},\bm{q};\bm{\tilde{\theta}}_P)$  will introduce lots of integer variables into the ICNN-Assisted OPF if we select ReLU functions as activation functions, thus making optimization problems intractable and NP-hard. Instead, we select a Softplus function as the activation function, which is  a smooth approximation of the ReLU function. The Softplus function does not introduce integer variables to express in the {ICNN-Assisted OPF}. In addition, as described in Proposition 3, the smooth property of Softplus function facilitates our algorithm design.

\begin{figure}
    \centering
    \includegraphics[width=3.2in]{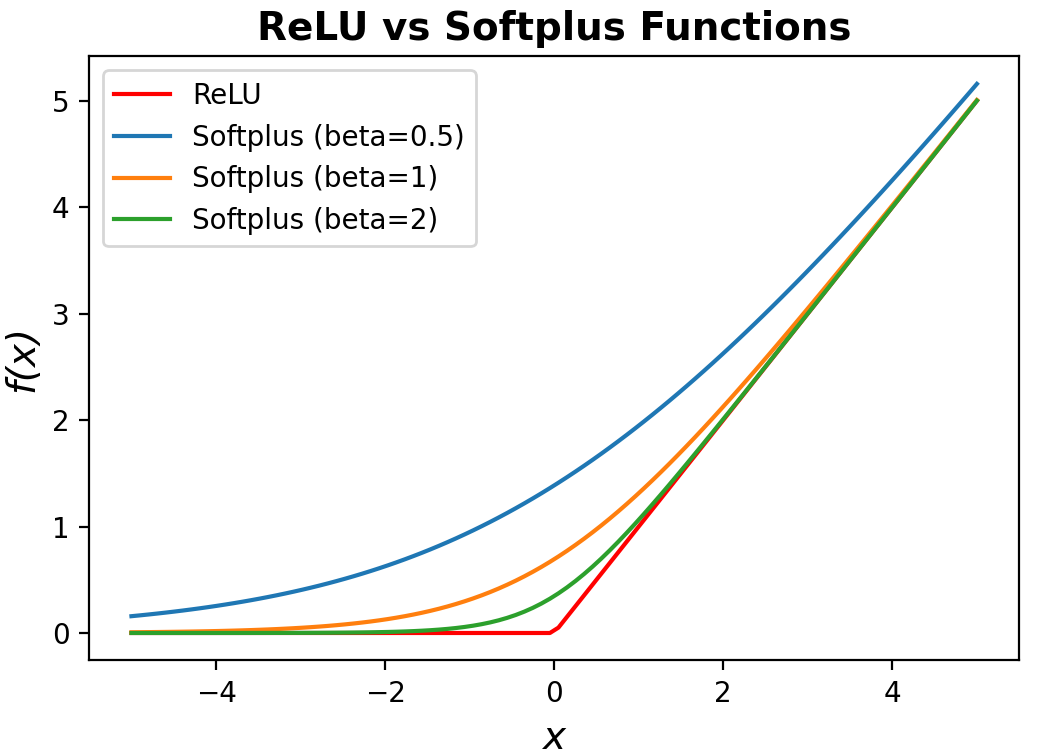}
    \caption{ReLU and Softplus function illustrations}
    \label{fig:softplus}
\end{figure}

More specifically, the Softplus function can be written as follows:
\begin{equation}
    \text{Softplus}(x)=\frac{1}{\beta}log(1+exp(\beta*x))
\end{equation}
As shown in Fig.\ref{fig:softplus}, the higher the value of $\beta$ is, the Softplus function is closer to the ReLU function. Customized Softplus functions can be designed by setting different $\beta$ values in deep learning frameworks, e.g., Pytorch.

Note that $f(\bm{p},\bm{q};\bm{\tilde{\theta}}_V)$ and $g(\bm{p},\bm{q};\bm{\tilde{\theta}}_P)$  could be extremely complex even though $f(\bm{p},\bm{q};\bm{\tilde{\theta}}_V)$ and $g(\bm{p},\bm{q};\bm{\tilde{\theta}}_P)$  do not introduce integer variables by using Softplus functions as activation functions. Most existing optimization solvers do not exhibit the capability to handle such high-dimension complex nonlinear constraints  (\ref{eq:ICNNOPF}b)-  (\ref{eq:ICNNOPF}c) well. Next subsection will discuss how to resolve this challenge. 
\subsection{ICNN-Assisted OPF Problem Reformulation}
For ease of notation, we define:
\begin{align*}
    \bm{\tilde{x}}&=
    \begin{bmatrix}
       \bm{p}^c\\
       \bm{q}^c
    \end{bmatrix},
    \bm{u}=\{\bm{p}^u,\bm{q}^u\}
    \\
    \mathcal{\tilde{X}}&=\{\bm{\tilde{x}}|\bm{p}^c\in\mathcal{P}^c,\bm{q}^c\in\mathcal{Q}^c\}
    \\ F(\bm{\tilde{x}};\bm{u},\bm{\tilde{\theta}}_V)&=f(\bm{p}^c+\bm{p}^u,\bm{q}^c+\bm{q}^{u};\bm{\tilde{\theta}}_V)\\
    G(\bm{\tilde{x}};\bm{u},\bm{\tilde{\theta}}_P)&=g(\bm{p}^c+\bm{p}^u,\bm{q}^c+\bm{q}^{u};\bm{\tilde{\theta}}_P)
\end{align*}
Motivated by \cite{YZC}, we expand the input vector from $\bm{\tilde{x}}$ to $\bm{{x}}$:
\begin{align*}
\bm{x}=\begin{bmatrix}
    \bm{\tilde{x}}\\
    -\bm{\tilde{x}}
\end{bmatrix}
\end{align*}
which includes both $\bm{\tilde{x}}$ and $-\bm{\tilde{x}}$. Considering $\bm{{x}}$ as the input, we utilize the ICNN framework to construct $F(\bm{{x}};\bm{u},\bm{{\theta}}_V)$ and $G(\bm{{x}};\bm{u},\bm{{\theta}}_P)$, learning the mapping from $\bm{{x}}$ to $\Big|\bm{V}-\frac{\bm{{V}}_{min}+ \bm{{V}}_{max}}{2}\Big|$ and $\Big|\bm{P}-\frac{\bm{{P}}_{min}+ \bm{{P}}_{max}}{2}\Big|$, respectively.

\medskip
\noindent
\textbf{[Proposition 2]} Using $\bm{x}$ as the ICNN input exhibits better mapping performance than using $\bm{\tilde{x}}$.

\noindent
\textbf{[Proof of Proposition 2]} For any given $\bm{\tilde{\theta}}_V$ and $\forall \bm{\tilde{x}}\in\mathcal{\tilde{X}}$, let $y=F(\bm{\tilde{x}};\bm{u},\bm{\tilde{\theta}}_V)$. Then, we set all weights in $\bm{{\theta}}_V$, related to $\bm{\tilde{x}}$ in $\bm{x}$, are the same as $\bm{\tilde{\theta}}_V$, and all weights in $\bm{{\theta}}_V$, related to $-\bm{\tilde{x}}$ in $\bm{x}$, are 0. It follows that $y=F(\bm{{x}};\bm{u},\bm{{\theta}}_V)$. Thus, we can conclude that for any given ICNN with the input $\bm{\tilde{x}}$, it is easy to find one ICNN with  the input $\bm{{x}}$ to achieve the same performance as the given ICNN with the input $\bm{\tilde{x}}$. The opposite is not true. The above analysis indicates the ICNN with the input $\bm{\tilde{x}}$ is a special case of ICNN with the input $\bm{{x}}$. Thus, using $\bm{x}$ as the ICNN input exhibits better mapping performance than using $\bm{\tilde{x}}$. Q.E.D.

We further define $\mathcal{X}$ as follows:
\begin{equation}
    \mathcal{X}=\big\{\bm{x}=
    [\bm{\tilde{x}}^T,\bm{w}^T]^T
    |\bm{\tilde{x}}\in\mathcal{\tilde{X}},
    \bm{w}=-\bm{\tilde{x}}
    \big\}
\end{equation}
From Proposition 2, we can conclude that $F(\bm{x};\bm{u},\bm{\theta}_V)$ and $G(\bm{x};\bm{u},\bm{\theta}_P)$ exhibit better approximation capabilities for $\Big|\bm{V}-\frac{\bm{{V}}_{min}+ \bm{{V}}_{max}}{2}\Big|$ and $\Big|\bm{P}-\frac{\bm{{P}}_{min}+ \bm{{P}}_{max}}{2}\Big|$ compared to $F(\bm{\tilde{x}};\bm{u},\bm{\tilde{\theta}}_V)$ and $G(\bm{\tilde{x}};\bm{u},\bm{\tilde{\theta}}_P)$. Thus, $F(\bm{x};\bm{u},\bm{\theta}_V)$ and $G(\bm{x};\bm{u},\bm{\theta}_P)$, instead of $F(\bm{\tilde{x}};\bm{u},\bm{\tilde{\theta}}_V)$ and $G(\bm{\tilde{x}};\bm{u},\bm{\tilde{\theta}}_P)$, are adopted in the ICNN-Assisted OPF, which can be rewritten as follows:
\begin{subequations}\label{eq:ProblemFormulation}
    \begin{align}
        \min_{\bm{x}\in\mathcal{X}} h(\bm{x})\\
        F(\bm{x};\bm{u},\bm{\theta}_V)\leq\Delta{\bm{V}}\\
        G(\bm{x};\bm{u},\bm{\theta}_P)\leq\Delta{\bm{P}}
    \end{align}
\end{subequations}

For this ICNN-Assisted OPF, we make the following assumption on it.

\noindent
[\textbf{Assumption}] The set $\mathcal{X}$ is closed, convex, and bounded. And the function $h(\bm{x})$ is continuously differentiable and convex. Define the gradient map as follows:
\begin{equation}
    H(\bm{x})=
    \nabla_{\bm{x}}h(\bm{x})
\end{equation}
Then, it is assumed that the gradient map $H(\bm{x})$ is Lipschitz continuous over the set $\mathcal{X}$ with constant $L_{H}$ , i.e.,
\begin{equation}
    ||H(\bm{x}_1)-H(\bm{x}_2)||_2\leq{L_{H}}||\bm{x}_1-\bm{x}_2||_2, \forall{\bm{x}_1,\bm{x}_2}\in\mathcal{X}
\end{equation}

\subsection{Fast Primal-Dual Gradient Method}

For our later purpose of algorithm design, we show the Lipschitz continuous properties of the gradients $\nabla_{\bm{x}}F(\bm{x};\bm{u},\bm{\theta}_V)$ and $\nabla_{\bm{x}}G(\bm{x};\bm{u},\bm{\theta}_P)$ in $\bm{x}$ in Proposition 3 when using the Softplus function as the activation function.

\medskip
\noindent
[\textbf{Proposition 3}] If the SoftPlus function is used as the activation function in the ICNN to express $F(\bm{x};\bm{u},\bm{\theta}_V)$ and $G(\bm{x};\bm{u},\bm{\theta}_P)$, then the gradients $\nabla_{\bm{x}}F(\bm{x};\bm{u},\bm{\theta}_V)$ and $\nabla_{\bm{x}}G(\bm{x};\bm{u},\bm{\theta}_P)$ are Lipschitz continuous in $\bm{x}$ over $\mathcal{X}$ with  constants $L_F$ and $L_G$, respectively. That is, for $\forall{\bm{x}_1,\bm{x}_2}\in\mathcal{X}$, we have:
\begin{subequations}\small
\begin{align}
    ||\nabla_FL(\bm{x}_1;\bm{u},\bm{\theta}_V)-\nabla_FL(\bm{x}_2;\bm{u},\bm{\theta}_V)||_2\leq{L_{F}}||\bm{x}_1-\bm{x}_2||_2\\
    ||\nabla_GL(\bm{x}_1;\bm{u},\bm{\theta}_P)-\nabla_GL(\bm{x}_2;\bm{u},\bm{\theta}_P)||_2\leq{L_{G}}||\bm{x}_1-\bm{x}_2||_2
\end{align}
\end{subequations}

\noindent
[\textbf{Proof of Proposition 3}] Without loss of generalization, assume a $\{m+1\}$-layer ICNN is utilized to learn the mapping from $\bm{x}$ to $F(\bm{x};\bm{u},\bm{\theta}_V)$, that is $F(\bm{x};\bm{u},\bm{\theta}_V)=\bm{W}_m^{(y)}\bm{y}_m+\bm{W}_m^{(x)}\bm{x}+\bm{b}_m$. 
\begin{equation}\label{eq:GF}
    \nabla_{\bm{x}}F(\bm{x};\bm{u},\bm{\theta}_V)=\bm{W}_m^{(y)}\nabla_{\bm{x}}\bm{y}_m+\bm{W}_m^{(x)}
\end{equation}

To prove that the gradient of $F(\bm{x};\bm{u},\bm{\theta}_V)$ is Lipschitz continuous in $\bm{x}$, we'll analyze how this Lipschitz continuous property propagates through the layers of the ICNN. For the initial layer, let $\bm{y}_1$ be:
\begin{equation}
\bm{y}_{1}=\delta_1\big(\bm{W}_0^{(y)}\bm{y}_0+\bm{W}_0^{(x)}\bm{x}+\bm{b}_0\big)
\end{equation}
The gradient of $\bm{y}_{1}$ with respect to $\bm{x}$ is as follows:
\begin{equation}
    \nabla_{\bm{x}}\bm{y}_{1}=\nabla\delta_1\big(\bm{W}_0^{(y)}\bm{y}_0+\bm{W}_0^{(x)}\bm{x}+\bm{b}_0\big)\bm{W}_0^{(x)}
\end{equation}
Note that the gradient of Softplus function is the sigmoid function, which is Lipschitz continuous with Lipschitz constant $L_s$ and bounded. It indicates $\nabla\delta_1$ is Lipschitz continuous and bounded as $\delta_1$ is a Softplus function. Given $\bm{W}_0^{(x)}$ is a constant matrix, it follows that $\nabla_{\bm{x}}\bm{y}_{1}$ is Lipschitz continuous and bounded. 

For $i=1,2,,...,m-1$, we have:
\begin{subequations}\label{eq:yi1}
\begin{align}
\bm{y}_{i+1}&=\delta_i\big(\bm{W}_i^{(y)}\bm{y}_i+\bm{W}_i^{(x)}\bm{x}+\bm{b}_i\big)\\
\nonumber\nabla_{\bm{x}}\bm{y}_{i+1}&=\nabla\delta_i\big(\bm{W}_i^{(y)}\bm{y}_i+\bm{W}_i^{(x)}\bm{x}+\bm{b}_i\big)\\
&\cdot(\bm{W}_i^{(y)}\nabla_{\bm{x}}\bm{y}_i+\bm{W}_i^{(x)})
\end{align} 
\end{subequations}
Assume $\nabla_{\bm{x}}\bm{y}_{i}$  is Lipschitz continuous with $L_i$ and bounded, we want to show that $\nabla_{\bm{x}}\bm{y}_{i+1}$  is Lipschitz continuous and bounded, for $\forall i=1,2,,...,m-1$. Since $\nabla\delta_i$ is Lipschitz continuous and bounded, and $\bm{W}_i^{(y)}\bm{y}_i+\bm{W}_i^{(x)}\bm{x}+\bm{b}_i$ is affine in $\bm{x}$, then the composition $\nabla\delta_i\big(\bm{W}_i^{(y)}\bm{y}_i+\bm{W}_i^{(x)}\bm{x}+\bm{b}_i\big)$ is Lipschitz continuous and bounded.  In addition, for any $\bm{x}_1,\bm{x}_2\in\mathcal{X}$, it follows that: 
\begin{equation}\label{eq:WL}
\begin{split}
&||(\bm{W}_i^{(y)}\nabla_{\bm{x}}\bm{y}_i|_{\bm{x}=\bm{x}_1}+\bm{W}_i^{(x)})\\
&-(\bm{W}_i^{(y)}\nabla_{\bm{x}}\bm{y}_i|_{\bm{x}=\bm{x}_2}+\bm{W}_i^{(x)})||\\
&=||\bm{W}_i^{(y)}(\nabla_{\bm{x}}\bm{y}_i|_{\bm{x}=\bm{x}_1}-\nabla_{\bm{x}}\bm{y}_i|_{\bm{x}=\bm{x}_2})||\\
&\leq
||\bm{W}_i^{(y)}||\cdot||(\nabla_{\bm{x}}\bm{y}_i|_{\bm{x}=\bm{x}_1}-\nabla_{\bm{x}}\bm{y}_i|_{\bm{x}=\bm{x}_2})||\\
&\leq||\bm{W}_i^{(y)}||L_i||\bm{x}_1-\bm{x}_2||
\end{split}
\end{equation}
The last inequality holds since $\nabla_{\bm{x}}\bm{y}_{i}$  is Lipschitz continuous with $L_i$. From (\ref{eq:WL}) and the bounded characteristic of $\nabla_{\bm{x}}\bm{y}_i$, we know $\bm{W}_i^{(y)}\nabla_{\bm{x}}\bm{y}_i+\bm{W}_i^{(x)}$ is Lipschitz continuous with constant $||\bm{W}_i^{(y)}||L_i$ and bounded.

For ease of notation, let $f_i$ and $g_i$ be:
\begin{subequations}
\begin{align}
    f_i(\bm{x})=\nabla\delta_i\big(\bm{W}_i^{(y)}\bm{y}_i+\bm{W}_i^{(x)}\bm{x}+\bm{b}_i\big)\\
    g_i(\bm{x})=\bm{W}_i^{(y)}\nabla_{\bm{x}}\bm{y}_i+\bm{W}_i^{(x)}
\end{align}
\end{subequations}
From the above analysis, we know both $f_i(\bm{x})$ and $g_i(\bm{x})$ are bounded and Lipschitz continuous. Let $L_{f_i}$ and $L_{g_i}$ denote the Lipschitz constants of $f_i(\bm{x})$ and $g_i(\bm{x})$, and $M_{f_i}$ and $M_{g_i}$
denote the bound of $||f_i(\bm{x})||$ and $||g_i(\bm{x})||$.
From (\ref{eq:yi1}b), we can express $\nabla_{\bm{x}}\bm{y}_{i+1}=f_i(\bm{x})g_i(\bm{x})$. For $\forall \bm{x}_1,\bm{x}_2\in\mathcal{X}$, we have:
\begin{equation}\label{eq:nb}\small
\begin{split}
    &||\nabla_{\bm{x}}\bm{y}_{i+1}|_{\bm{x}=\bm{x}_1}-\nabla_{\bm{x}}\bm{y}_{i+1}|_{\bm{x}=\bm{x}_2}||\\&=||
    f_i(\bm{x}_1)g_i(\bm{x}_1)-f_i(\bm{x}_2)g_i(\bm{x}_2)
    ||\\
    &=||
    f_i(\bm{x}_1)(g_i(\bm{x}_1)-g_i(\bm{x}_2))+(f_i(\bm{x}_1)-f_i(\bm{x}_2))g_i(\bm{x}_2)
    ||\\
    &\leq
    ||
    f_i(\bm{x}_1)(g_i(\bm{x}_1)-g_i(\bm{x}_2))||+
    ||(f_i(\bm{x}_1)-f_i(\bm{x}_2))g_i(\bm{x}_2)||
    \\
    &\leq|| f_i(\bm{x}_1)||\cdot||g_i(\bm{x}_1)-g_i(\bm{x}_2)||+
    || g_i(\bm{x}_2)||\cdot||f_i(\bm{x}_1)-f_i(\bm{x}_2)||\\
    &\leq\Big[|| f_i(\bm{x}_1)||L_{g_i}+|| g_i(\bm{x}_2)||L_{f_i}\Big]||\bm{x}_1-\bm{x}_2||\\
    &\leq\Big(
    M_{f_i}
    L_{g_i}+M_{g_i}L_{f_i}\Big)
    ||\bm{x}_1-\bm{x}_2||
\end{split}
\end{equation}
It follows $\nabla_{\bm{x}}\bm{y}_{i+1}$  is Lipschitz continuous and bounded. Next, we can conclude that $\nabla_{\bm{x}}\bm{y}_{m}$  is Lipschitz continuous and bounded. Together with (\ref{eq:GF}), it follows that:
\begin{equation}
\begin{split}
    ||\nabla_{\bm{x}}F(\bm{x};\bm{u},\bm{\theta}_V)|_{\bm{x}=\bm{x}_1}-\nabla_{\bm{x}}F(\bm{x};\bm{u},\bm{\theta}_V)|_{\bm{x}=\bm{x}_2}||\\
    \leq
    ||\bm{W}_m^{(y)}||\cdot||\nabla_{\bm{x}}\bm{y}_{i+1}|_{\bm{x}=\bm{x}_1}-\nabla_{\bm{x}}\bm{y}_{i+1}|_{\bm{x}=\bm{x}_2}||\\
    \leq
    ||\bm{W}_m^{(y)}||\cdot\Big(
    M_{f_i}
    L_{g_i}+M_{g_i}L_{f_i}\Big)||\bm{x}_1-\bm{x}_2||
\end{split}
\end{equation}
Thus, $\nabla_{\bm{x}}F(\bm{x};\bm{u},\bm{\theta}_V)$ is Lipschitz continuous with constant $L_F=||\bm{W}_m^{(y)}||\cdot\Big(
    M_{f_i}
    L_{g_i}+M_{g_i}L_{f_i}\Big)$.
The Lipschitz continuous of $\nabla_{\bm{x}}G(\bm{x};\bm{u},\bm{\theta}_V)$ can be proved in the same way. Q.E.D.

Next, consider the Lagrangian function $L(\bm{x},\bm{\lambda}_V,\bm{\lambda}_P)$ for (\ref{eq:ProblemFormulation}) as follows:
\begin{equation}
\begin{split}
L(\bm{x},\bm{\lambda}_V,\bm{\lambda}_P)&=h(\bm{x})+\bm{\lambda}^T_V\Big[F(\bm{x};\bm{u},\bm{\theta}_V)-\Delta{\bm{V}}\Big]\\
&+\bm{\lambda}^T_P\Big[G(\bm{x};\bm{u},\bm{\theta}_P)-\Delta{\bm{P}}\Big]
\end{split}
\end{equation}
In lieu of $L(\bm{x},\bm{\lambda}_V,\bm{\lambda}_P)$, introduce the following regularized Lagrangian function:
\begin{equation}\label{eq:RLang}
\begin{split}
    L_{\upsilon,\epsilon}(\bm{x},\bm{\lambda}_V,\bm{\lambda}_P)&=L(\bm{x},\bm{\lambda}_V,\bm{\lambda}_P)+\frac{\upsilon}{2}||\bm{x}||_2^2\\
    &-\frac{\epsilon}{2}(||\bm{\lambda}_V||_2^2+||\bm{\lambda}_P||_2^2)
\end{split}
\end{equation}
where the constant $\upsilon>0$ and $\epsilon>0$ are design parameters. Function (\ref{eq:RLang}) is strictly convex in $\bm{x}$ and strictly concave in the dual variables $\bm{\lambda}_V,\bm{\lambda}_P$. Moreover, consider the regularized mapping $\bm{\Phi}(\bm{x},\bm{\lambda}_V,\bm{\lambda}_P)$ as follows:
\begin{equation}
    \bm{\Phi}(\bm{x},\bm{\lambda}_V,\bm{\lambda}_P)=
    \begin{bmatrix}
        \nabla_{\bm{x}}{L}_{\upsilon,\epsilon}(\bm{x},\bm{\lambda}_V,\bm{\lambda}_P)\\
        \nabla_{\bm{\lambda}_V}{L}_{\upsilon,\epsilon}(\bm{x},\bm{\lambda}_V,\bm{\lambda}_P)\\
        \nabla_{\bm{\lambda}_P}{L}_{\upsilon,\epsilon}(\bm{x},\bm{\lambda}_V,\bm{\lambda}_P)]
    \end{bmatrix}
\end{equation}
\begin{subequations}\label{eq:Lgradient}
\begin{align}
    \nonumber\nabla_{\bm{x}}{L}_{\upsilon,\epsilon}(\bm{x},\bm{\lambda}_V,\bm{\lambda}_P)&=
    \nabla_x\bm{h}(\bm{x})
    +
    [\nabla_{\bm{x}}{F}(\bm{x};\bm{u},\bm{\theta}_V)]^T\bm{\lambda}_V\\
    &+
    [\nabla_{\bm{x}}{G}(\bm{x};\bm{u},\bm{\theta}_P)]^T\bm{\lambda}_P
    +\upsilon\bm{x}
    \\
    \nabla_{\bm{\lambda}_V}{L}_{\upsilon,\epsilon}(\bm{x},\bm{\lambda}_V,\bm{\lambda}_P)&=F(\bm{x};\bm{u},\bm{\theta}_V)-\Delta{\bm{V}}-\epsilon\bm{\lambda}_V\\
     \nabla_{\bm{\lambda}_P}{L}_{\upsilon,\epsilon}(\bm{x},\bm{\lambda}_V,\bm{\lambda}_P)&=G(\bm{x};\bm{u},\bm{\theta}_P)-\Delta{\bm{P}}-\epsilon\bm{\lambda}_P
\end{align}
\end{subequations}
where $\nabla_{\bm{x}}{F}(\bm{x};\bm{u},\bm{\theta}_V)$ and $\nabla_{\bm{x}}{G}(\bm{x};\bm{u},\bm{\theta}_V)$ are the Jacobian matrices  of $F$ and $G$ with respect to $\bm{x}$.

\medskip
\noindent
[\textbf{Lemma}] As \textbf{Assumption} holds, then  $\bm{\Phi}(\bm{x},\bm{\lambda}_V,\bm{\lambda}_P)$ is strongly monotone over $\mathcal{X}\times\mathbb{R}_+\times\mathbb{R}_+$ with constant $\eta=\min\{\upsilon,\epsilon\}$ and Lipschitz continuous over $\mathcal{X}\times\mathbb{R}_+\times\mathbb{R}_+$ with constant $L_{\Phi}$. 

 The proof of [\textbf{Lemma}] is provided in \cite[Lemma 3.4]{JKA}. The upshot of (\ref{eq:RLang}) is that gradient-based methods can be applied to find an approximate solution of the ICNN-Assisted OPF (\ref{eq:ProblemFormulation}) with improved convergence properties \cite{JKA,ASG}. More specifically, it aims to construct the approximate solution of (\ref{eq:ProblemFormulation}) by solving the following approximate saddle-point problem:
\begin{equation}\label{eq:RLangOPT}
    \max_{\bm{\lambda}_V\in\mathbb{R}_+,\bm{\lambda}_P\in\mathbb{R}_+}\min_{\bm{x}\in\mathcal{X}}L_{\upsilon,\epsilon}(\bm{x},\bm{\lambda}_V,\bm{\lambda}_P)
\end{equation}
whose  unique optimal primal-dual optimizer is $\bm{z}^*=\{\bm{x}^*,\bm{\lambda}_V^*,\bm{\lambda}_P^*\}$. 

Here, consider the following primal-dual gradient method to solve the problem (\ref{eq:RLangOPT}) to obtain $\bm{z}^*$ as follows:

\noindent
\textbf{Primal-Dual Gradient Method:}
\\
For any iteration $k$, we update:
\begin{subequations}\label{eq:primaldual}
    \begin{align}
    \bm{x}^{k+1}&=\text{Proj}_{\mathcal{X}}
    \Big\{\bm{x}^k-\mu\nabla_{\bm{x}}{L}_{\upsilon,\epsilon}(\bm{x}^k,\bm{\lambda}_V^k,\bm{\lambda}_P^k)\Big\}
       \\
       \bm{\lambda}_V^{k+1}&=\text{Proj}_{\mathbb{R}_{+}}\Big\{\bm{\lambda}_V^{k}+\mu\nabla_{\bm{\lambda}_V}{L}_{\upsilon,\epsilon}(\bm{x}^k,\bm{\lambda}_V^k,\bm{\lambda}_P^k)\Big\}\\
    \bm{\lambda}_P^{k+1}&=\text{Proj}_{\mathbb{R}_{+}}\Big\{\bm{\lambda}_P^{k}+\mu\nabla_{\bm{\lambda}_P}{L}_{\upsilon,\epsilon}(\bm{x}^k,\bm{\lambda}_V^k,\bm{\lambda}_P^k)\Big\} 
    \end{align}
\end{subequations}
where $\mu>0$ is the step size.  Next,  we present the convergence of sequence $\{\bm{z}^{k}\}$ with $\bm{z}^k=\{\bm{x}^k,\bm{\lambda}_V^k,\bm{\lambda}_P^k\}$ generated by (\ref{eq:primaldual}).

\subsection{Theoretical Analysis}
\noindent
[\textbf{Proposition 4}] Let Assumption and Proposition 3 hold. If the step size $\mu>0$ is chosen such that:
\begin{equation}
    \rho({\mu})=\sqrt{1-2\mu\eta+\mu^2{L}_{\Phi}^2}<1
\end{equation}
that is $\mu\leq\frac{2\eta}{L_{\Phi}^2}$, then the sequence $\{\bm{z}^k\}=\{\bm{x}^k,\bm{\lambda}_V^k,\bm{\lambda}_P^k\}$, determined by (\ref{eq:primaldual}), converges to $\bm{z}^*$:
\begin{equation}
    \lim_{k\rightarrow\infty}||\bm{z}_{k}-\bm{z}^*||_2=0
\end{equation}

\noindent
[\textbf{Proof of Proposition 4}]
The update (\ref{eq:primaldual}) of primal-dual gradient method can be equivalently expressed as follows:
\begin{equation}    \bm{z}^{k+1}=\text{Proj}_{\mathcal{X}\times\mathbb{R}_{+}\times\mathbb{R}_{+}}\{\bm{z}^k-\mu\bm{\Phi}(\bm{z}^k)\}
\end{equation}
By standard optimality conditions, the optimizer is a fixed point of the iterations, i.e., $\bm{z}^{*}=\text{Proj}_{\mathcal{X}\times\mathbb{R}_{+}\times\mathbb{R}_{+}}\{\bm{z}^*-\mu\bm{\Phi}(\bm{z}^*)\}$, then we have :
\begin{equation}\small
\begin{split}
    ||\bm{z}^{k}-\bm{z}^*||_2=&||\text{Proj}_{\mathcal{X}\times\mathbb{R}_{+}\times\mathbb{R}_{+}}\big\{\bm{z}^{k-1}-\mu\bm{\Phi}(\bm{z}^{k-1})\big\}-\bm{z}^*
    ||_2\\
    =&||\text{Proj}_{\mathcal{X}\times\mathbb{R}_{+}\times\mathbb{R}_{+}}\big\{\bm{z}^{k-1}-\mu\bm{\Phi}(\bm{z}^{k-1})\big\}-
    \\
    &\text{Proj}_{\mathcal{X}\times\mathbb{R}_{+}\times\mathbb{R}_{+}}
    \big\{\bm{z}^{*}-\mu\bm{\Phi}(\bm{z}^{*})
    \big\}
    ||_2
\end{split}
\end{equation}
We  utilize the non-expansivity property of the projection
operator, which yields:
\begin{equation}\label{eq:norm2}
\begin{split}
    ||\bm{z}^{k}-\bm{z}^*||_2\leq&||\bm{z}^{k-1}-\mu\bm{\Phi}(\bm{z}^{k-1})-
    \\
    &
   \bm{z}^{*}+\mu\bm{\Phi}(\bm{x}^{*})
    ||_2  
\end{split}
\end{equation}
From [\textbf{Lemma}], it follows that:
\begin{equation}\label{eq:1}
\begin{split}
    &(\bm{z}^{k-1}-\bm{z}^*)^T(\bm{\Phi}(\bm{z}^{k-1})-\bm{\Phi}(\bm{z}^{*}))\\&\geq\eta||\bm{z}^{k-1}-\bm{z}^*||_2^2||\bm{z}^{k-1}-\bm{z}^{*}||_2^2
\end{split}
\end{equation}
\begin{equation}\label{eq:2}
    \begin{split}
        ||\bm{\Phi}(\bm{z}^{k-1})-\bm{\Phi}(\bm{z}^{*})||_2^2\leq{L}_{\Phi}^2||\bm{z}^{k-1}-\bm{z}^{*}||_2^2
    \end{split}
\end{equation}
From (\ref{eq:norm2}), we have:
\begin{equation}\label{eq:3}\small
\begin{split}
    &||\bm{z}^{k-1}-\mu\bm{\Phi}(\bm{z}^{k-1})-\bm{z}^{*}+\mu\bm{\Phi}(\bm{z}^{*})
    ||_2^2=  
    \\&||\bm{z}^{k-1}-\bm{z}^*||_2^2+\mu^2||\bm{\Phi}(\bm{z}^{k-1})-\bm{\Phi}(\bm{z}^{*})||_2^2\\
    &-2\mu(\bm{z}^{k-1}-\bm{z}^*)^T(\bm{\Phi}(\bm{z}^{k-1})-\bm{\Phi}(\bm{z}^{*}))
\end{split}
\end{equation}
By putting together the results in (\ref{eq:1}),(\ref{eq:2}), and (\ref{eq:3}), we have:
\begin{equation}
    \begin{split}
        ||\bm{z}^{k}-\bm{z}^*||_2^2\leq(1-2\mu\eta+\mu^2{L}_{\Phi}^2)||\bm{z}^{k-1}-\bm{z}^{*}||_2^2
    \end{split}
\end{equation}
\begin{equation}\label{eq:znorm}
    \begin{split}
        ||\bm{z}^{k}-\bm{z}^*||_2\leq\sqrt{1-2\mu\eta+\mu^2{L}_{\Phi}^2}||\bm{z}^{k-1}-\bm{z}^{*}||_2
    \end{split}
\end{equation}
As $\rho({\mu})<1$, i.e., $\mu\leq\frac{2\eta}{L_{\Phi}^2}$, then (\ref{eq:znorm}) represents a contraction. It follows that:
\begin{equation}
    ||\bm{z}_{k}-\bm{z}^*||_2\leq[\rho({\mu})]^k||\bm{z}^{0}-\bm{z}_{*}||_2
\end{equation}
\begin{equation}
    \lim_{k\rightarrow\infty}||\bm{z}^{k}-\bm{z}^*||_2=0
\end{equation}
Q.E.D.

\noindent
[\textbf{Remark 2}] 
From  (\ref{eq:Lgradient}) and (\ref{eq:primaldual}), the key to performing  the primal-dual gradient method is to calculate $\nabla_{\bm{x}}{F}(\bm{x};\bm{u},\bm{\theta}_V)$ and $\nabla_{\bm{x}}{G}(\bm{x};\bm{u},\bm{\theta}_P)$. Note that both $\nabla_{\bm{x}}{F}(\bm{x};\bm{u},\bm{\theta}_V)$ and $\nabla_{\bm{x}}{G}(\bm{x};\bm{u},\bm{\theta}_P)$ are essentially the gradient of the ICNN outputs with respect to its input, which can be easily and efficiently calculated by using \textit{the chain rule} in the ICNN. Such gradients of ICNN can also be directly accessed by open-source machine learning frameworks, e.g., Pytorch. Utilizing \textit{the chain rule}  can efficiently speed up the implementation of  gradient calculation.

\section{ICNN-Assisted OPF Applications}
\label{sec:application}
The ICNN-Assisted OPF can be applied to different distribution network structures and applications by choosing different objective functions and data sources to learn  ${F}(\bm{x};\bm{u},\bm{\theta}_V)$ and ${G}(\bm{x};\bm{u},\bm{\theta}_P)$. Without loss of generality, possible distribution network applications by utilizing our proposed ICNN-Assisted OPF are as follows:

(1) Volt/VAr Optimization (VVO) Problem: VVO plays a key role in the operation and management of distribution networks. It involves the control and optimization of voltage levels (Volt) and reactive power (VAr) within the grid to achieve multiple objectives, such as minimizing voltage deviations, reducing energy losses, and maintaining system reliability. For the VVO problem, the decision variable is the reactive power vector $\bm{q}^c$.

(2) Distributed Energy Resource (DER) Management Problem: DER management problems aim to  achieve different goals, e.g.,  social welfare maximization, cost minimization, by dispatching and controlling DERs, subject individual DER and distribution network constraints. For example, the goal of DER management optimization problems can be maximizing the net benefit of all DERs, then $h(\bm{x})$ with $\bm{x}=\{\bm{p}^c,\bm{q}^c\}$ can be set as:
\begin{equation}
    h(\bm{x})=\sum_{i\in\mathcal{N}} \big[U_i(p_i^c)-C_i(p_i^c)\big]
\end{equation}
where $U_i(p_i^c)$ denotes the utility agent $i$ obtains by using $p_i^c$, and  $C_i(p_i^c)$ denotes the cost agent $i$ pays for $p_i^c$.

(3) Coordinated Real and Reactive Power Optimization Problem: This problem involves the optimal management and control of both real (active) and reactive power to ensure efficient, reliable, and stable operation of the electrical distribution system. Taking photovoltaic (PV) as an example, we can simultaneously shed the PV real power and control the PV reactive power to protect distribution networks from violations of network constraints. In this case, the ICNN-Assisted OPF can be expressed by:
\begin{subequations}\label{eq:PQOPF}
\begin{align}
    \min_{\bm{x}} &~h(\bm{x})=\sum_{i\in\mathcal{N}}\big[C_i^p(p_i^c)+C_i^q(q_i^c)\big]\\
    \text{subject to:}~~
    & f(\bm{p},\bm{q};\bm{\theta}_V)\leq\Delta{\bm{V}}\\
    &g(\bm{p},\bm{q};\bm{\theta}_P)\leq\Delta{\bm{P}}\\
    & \bm{p}=\bm{p}^c+\bm{p}^u\\
    & \bm{q}=\bm{q}^c+\bm{q}^u\\
    &\bm{p}^c\leq\bm{\overline{p}}^c\\
    &({p}_i^c)^2+({q}_i^c)^2\leq (\overline{S}_{i})^2, \forall i \in\mathcal{N}
\end{align}
\end{subequations}
where $C_i(p_i^c)$ and $C_i(q_i^c)$ denote the cost generated by utilizing  $p_i^c$ and $q_i^c$, $\bm{\overline{p}}^c$ is the maximum PV real power output, and $\overline{S}_{i}$ is the capacity of $i$-th PV inverter.

In addition, the ICNN-Assisted OPF can also be applied to other research areas and topics, including economic dispatch, EV charging schedule, and so on.
\section{Numerical Case Study}
\label{sec:case}
\subsection{Experimental Setup}
\begin{figure}
    \centering
    \includegraphics[width=3.5in]{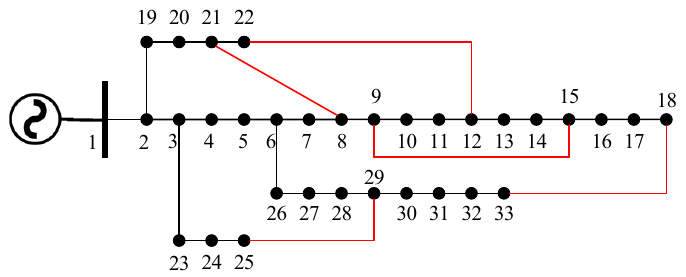}
    \caption{Modified meshed IEEE 33 bus system}
    \label{fig:IEEE33}
\end{figure}

\begin{figure}
    \centering
    \includegraphics[width=3.3in]{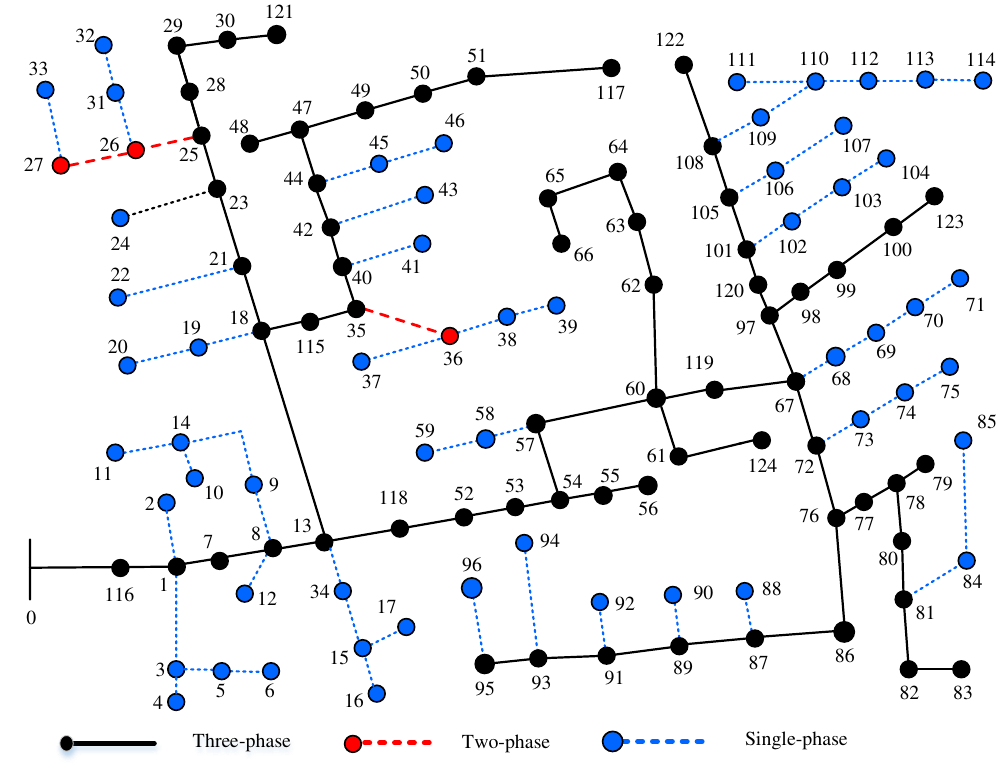}
    \caption{Unbalanced IEEE 123 bus system}
    \label{fig:IEEE123}
\end{figure}

We evaluate the performance of our proposed ICNN-Assisted OPF on the modified meshed IEEE 33 bus system, shown in Fig.\ref{fig:IEEE33}, and the unbalanced IEEE 123 bus system, shown in Fig.\ref{fig:IEEE123}, in this section, further testing the ICNN-Assisted OPF on distribution networks of different structures. The base power and voltage for the modified meshed IEEE 33 bus system are set as: $S_{\sf base}=100$ (kVA) and $V_{\sf base}$ = $4.16$ (kV), and the base power and voltage  for the unbalanced IEEE 123 bus system are set as: $S_{\sf base}=100$ (kVA) and $V_{\sf base}$ = $12.66$ (kV). All these simulations are performed using the Pytorch framework. 


\subsection{ICNN Accuracy Verification}
To verify the accuracy of ICNN, we compare the power flow solutions, calculated by the ICNN, with other power flow algorithms. Different power flow models are considered as follows:

\textbf{[A1]. Nonlinear Power Flow Model}: It utilizes the balanced and unbalanced nonlinear power flows for single-phase and three-phase distribution networks, which can be regarded as {the benchmark}.  The balanced and unbalanced nonlinear power flows are calculated by Matpower \cite{MATPOWER} and open-source Open Distribution System Simulator (OpenDSS) \cite{OpenDSS}, respectively.

\textbf{[A2]. Linearized Power Flow Model}: It makes full use of LinDistFlow \cite{MB}-\cite{RCZ}, where linearized balanced and unbalanced  power flows are applied to single-phase and three-phase distribution networks, respectively. \textit{Note that the linearized balanced and unbalanced power flows are not suitable to be applied in meshed distribution networks.}

\textbf{[A3]. Neural Network Model}: Neural networks are leveraged to learn the mapping from control variables to power flow solutions, e.g., voltage magnitudes, line power flows. Here, we construct standard 4-layer neural networks.

\textbf{[A4]. Proposed ICNN Model}: ICNN with SoftPlus activation functions is considered to learn the mapping from control variables to power flow solutions, e.g., voltage magnitudes, line power flows. We also construct the standard 4-layer ICNN and keep the number of layer and matrices the same dimension as those of neural networks.

\begin{table}[h]
\normalsize
\centering
\caption{{MSE Comparisons among different models in the meshed IEEE 33 bus system under 500 randomly generated scenarios}}
\label{tab:MSE33}
\footnotesize
\renewcommand\arraystretch{1.0}
        \begin{tabular}{cccccc}
        \hline
        \hline
          Model& $\textbf{A2}$ & $\textbf{A3}$ & $\textbf{A4}$ \\
        \hline
        Voltage Deviation MSE& --&1.335e-7&1.291e-6\\
        Line Flow Deviation MSE& --&1.447e-4&1.812e-4\\
        \hline
        \hline
        \end{tabular}
\end{table}

\begin{table}[h]
\normalsize
\centering
\caption{{MSE Comparisons among different models in the unbalanced IEEE 123 bus system under 500 randomly generated scenarios}}\label{tab:MSE123}
\footnotesize
\renewcommand\arraystretch{1.0}
        \begin{tabular}{cccccc}
        \hline
        \hline
          Model& $\textbf{A2}$ & $\textbf{A3}$ & $\textbf{A4}$\\
        \hline
        Voltage Deviation MSE& 1.615e-4&3.261e-7&8.455e-7\\
        Line Flow Deviation MSE& 8.993e-3&1.761e-3&4.632e-3\\
        \hline
        \hline
        \end{tabular}
\end{table}

\begin{figure}[htb]
    \centering
    \includegraphics[width=3.2in]{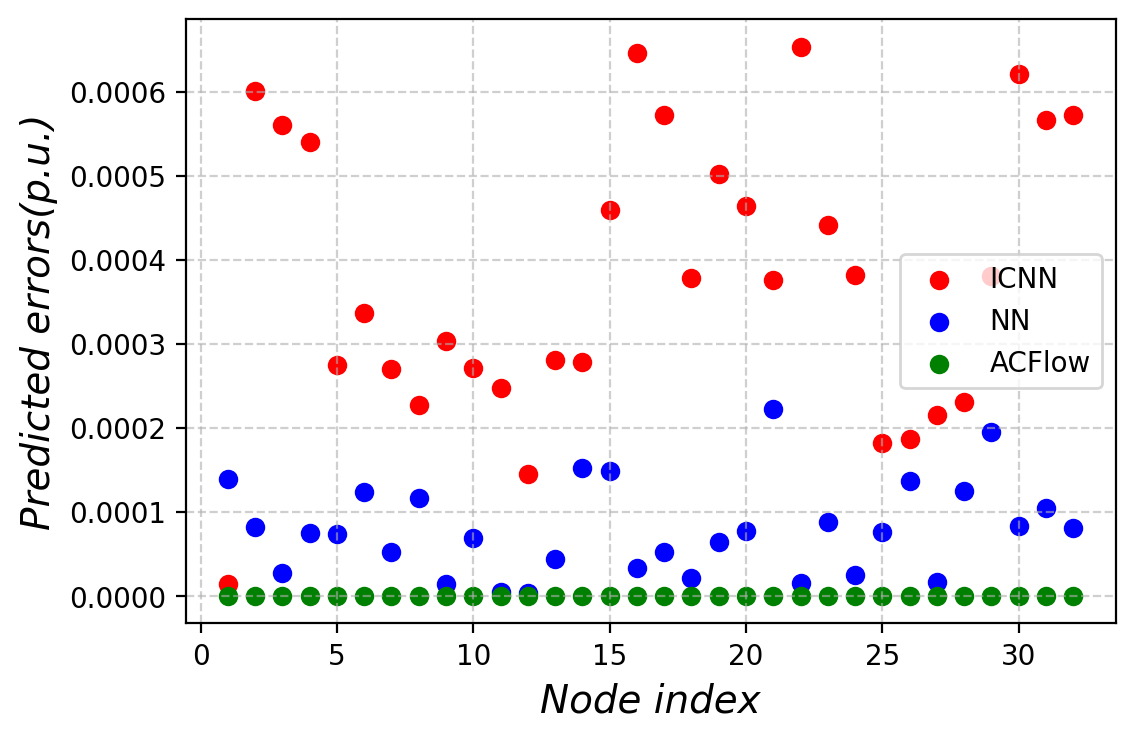}
    \caption{
    Comparisons of prediction errors on nodal voltage deviations for the modified meshed IEEE 33 bus system under one scenario}
    \label{fig:VDeviation33}
\end{figure}

\begin{figure}[htb]
    \centering
    \includegraphics[width=3.2in]{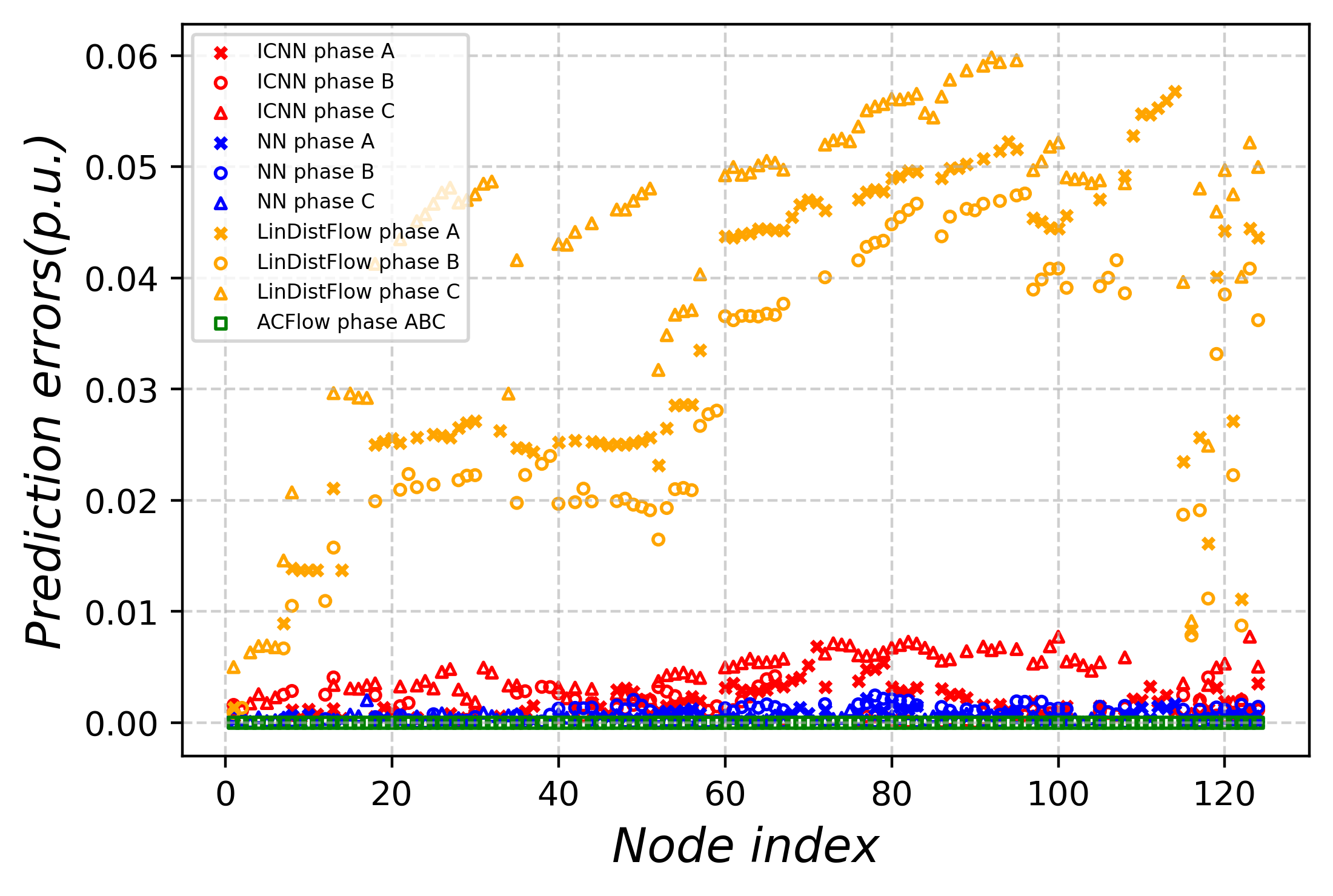}
    \caption{Comparisons of prediction errors on nodal voltage deviations for the unbalanced IEEE 123 bus system under one random scenario}
    \label{fig:VDeviation123}
\end{figure}

We test the performance of \textbf{A1-A4} on different types of distribution network, including a modified meshed IEEE {33 bus} system, an unbalanced three-phase IEEE 123 bus system. The mean squared error (MSE) of $\Big|\bm{V}-\frac{\bm{{V}}_{min}+ \bm{{V}}_{max}}{2}\Big|$ and $\Big|\bm{P}-\frac{\bm{{P}}_{min}+ \bm{{P}}_{max}}{2}\Big|$ under 500 randomly generated scenarios are given in Tab.\ref{tab:MSE33} and Tab.\ref{tab:MSE123}, where \textbf{A1} is regarded as the benchmark. As depicted in Tab.\ref{tab:MSE33} and Tab.\ref{tab:MSE123}, \textbf{A3} exhibits the lowest MSE compared to \textbf{A2} and \textbf{A4}, indicating the neural network model exhibits the best capability to approximate the nonlinear power flow \textbf{A1}. In addition, \textbf{A4} also shows a great capability to follow the nonlinear power flow \textbf{A1}, slight worse than \textbf{A3} but much better than \textbf{A2}. The neural network model is powerful in mapping non-convex and nonlinear functions, leading to the lowest MSE. The  generalization ability of our proposed ICNN model is slightly inferior to the neural network due to its convex mapping design, but the ICNN model is able to express the nonlinear properties. Instead, the linearized power flow model fails to exhibit the nonlinear properties. Thus, the performance of  \textbf{A3} is much better than \textbf{A2}. The  prediction errors of other models on voltage deviations, i.e., \textbf{A2}-\textbf{A4}, compared to \textbf{A1} are depicted in Fig.\ref{fig:VDeviation33} and Fig.\ref{fig:VDeviation123}, respectively.

\begin{figure}[htb]
    \centering
    \includegraphics[width=3.2in]{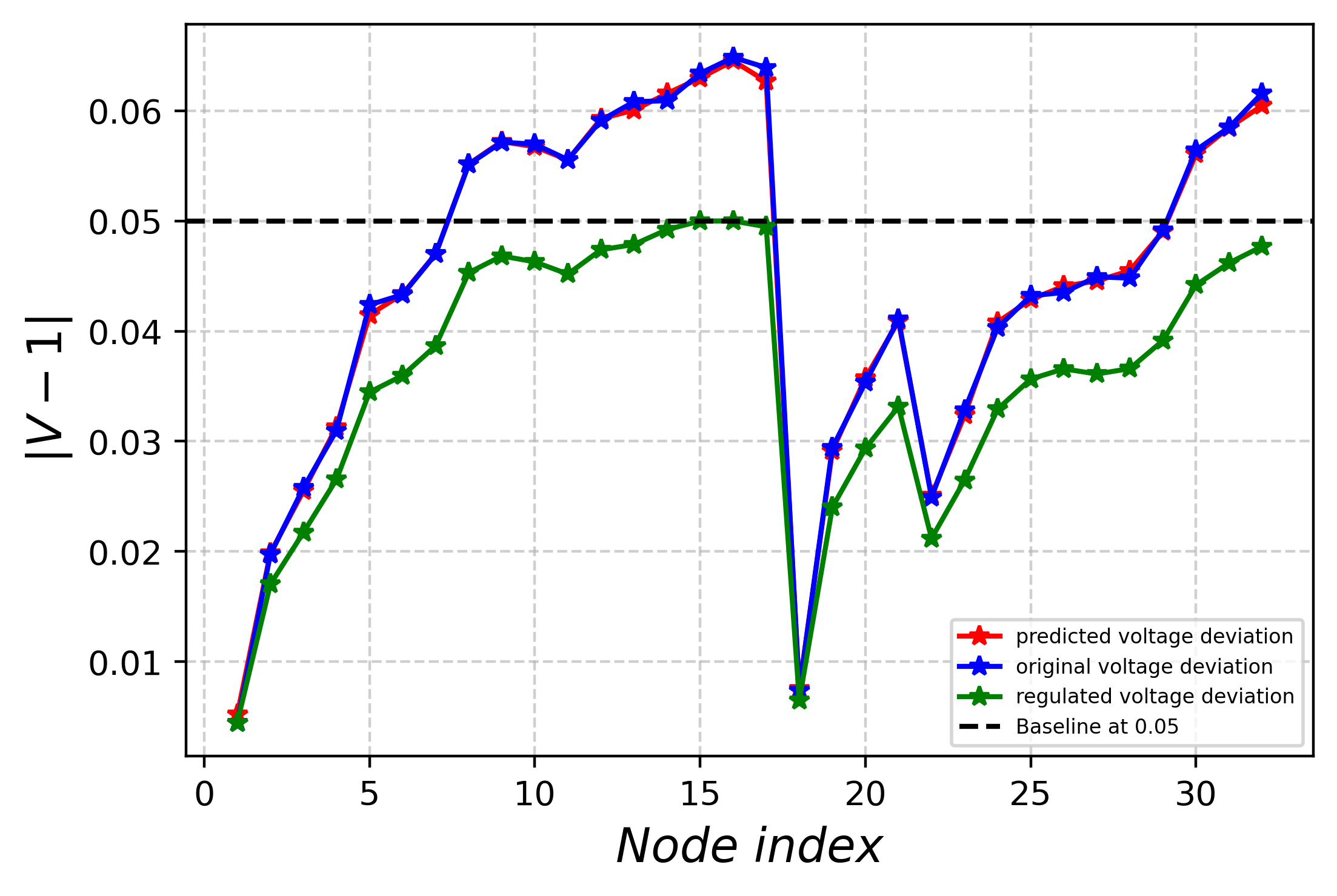}
    \caption{The absolute voltage deviation across the modified meshed IEEE 33 bus system: the blue color denotes the original voltages without any control; the red color denotes the predicted voltages using ICNN; the green color denotes the regulated voltage deviations by means of ICNN-Assisted OPF}
    \label{fig:NetworkState33}
\end{figure}

\subsection{Distribution Network OPF Performance}
Taking the coordinated real and reactive power optimization problem (\ref{eq:PQOPF}) as an example, its goal is to ensure the safe and reliable operation of distribution networks with  minimal  costs. Here, we  test the performance of ICNN-Assisted OPF on  the modified meshed IEEE {33 bus} system and the unbalanced three-phase IEEE 123 bus system.

\begin{figure}
    \centering
    \includegraphics[width=3.5in]{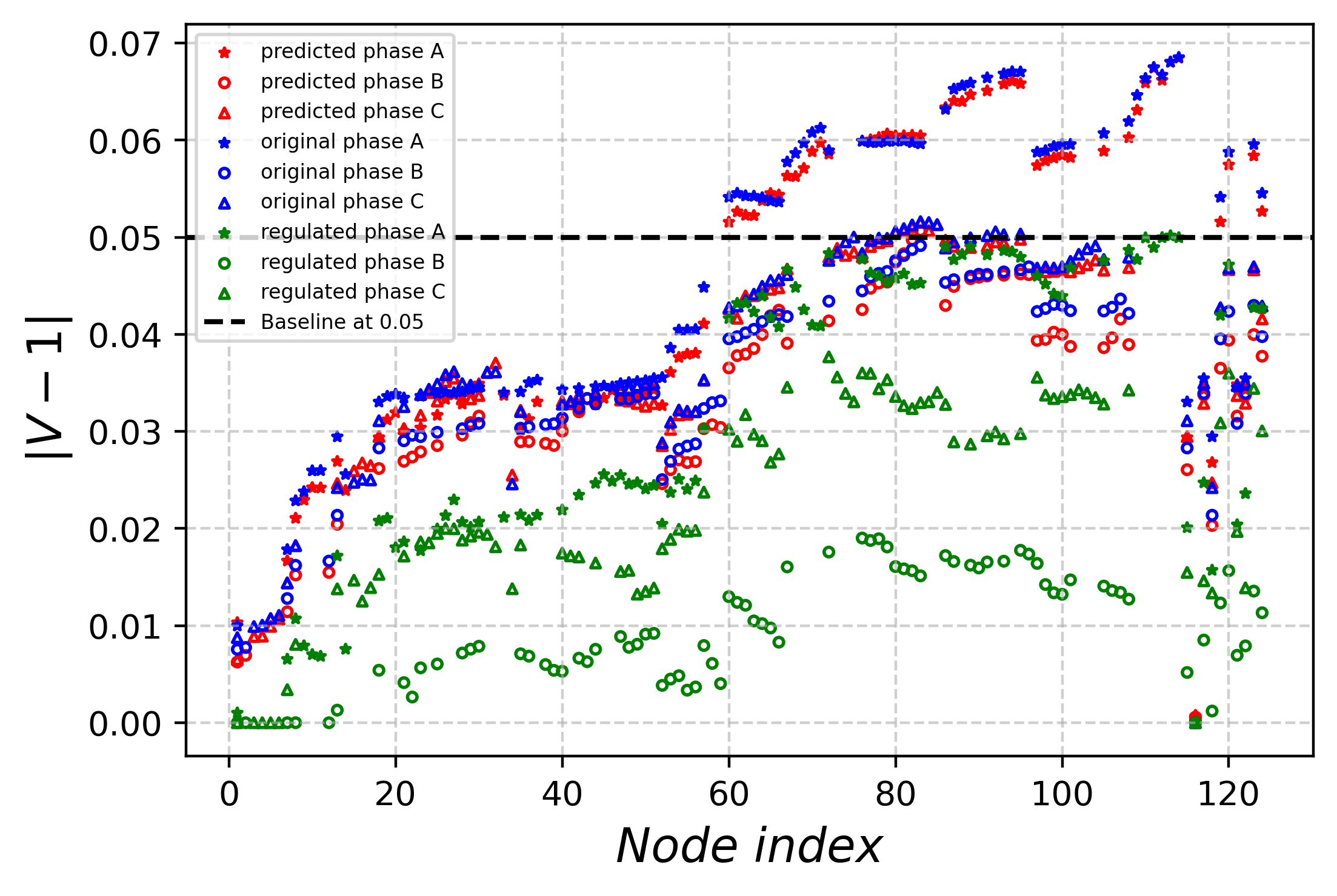}
    \caption{The absolute voltage deviation across the unbalanced IEEE 123 bus system: the blue color denotes the original voltages without any control; the red color denotes the predicted voltages using ICNN; the green color denotes the regulated voltage deviations by means of ICNN-Assisted OPF}
    \label{fig:NetworkState}
\end{figure}

In the  modified meshed  IEEE {33 bus} system, we  set $C_i^p(p_i^c)=\frac{1}{2}{p}_i^c{d}_i^p{p}_i^c$ and  $C_i^q(q_i^c)=\frac{1}{2}{q}_i^c{d}_i^q{q}_i^c$ with cost parameters  ${d}_i^p$ and ${d}_i^q$, and  utilize the operation data of meshed IEEE 33 bus system to learn (\ref{eq:PQOPF}b)-(\ref{eq:PQOPF}c).  There is no any violation of voltage or power flow by means of ICNN-Assisted OPF in the modified meshed IEEE 33 bus system. For example, the absolute voltage deviation results of modified meshed IEEE 33 bus system are depicted in Fig.\ref{fig:NetworkState33}. As shown in Fig.\ref{fig:NetworkState33}, the ICNN can achieve a good prediction for the voltage deviations in the meshed IEEE 33 bus system as there is no any control, and the ICNN-Assisted OPF can also protect the  meshed IEEE 33 bus system from voltage constraint violations.

\begin{figure}
    \centering
    \includegraphics[width=3.5in]{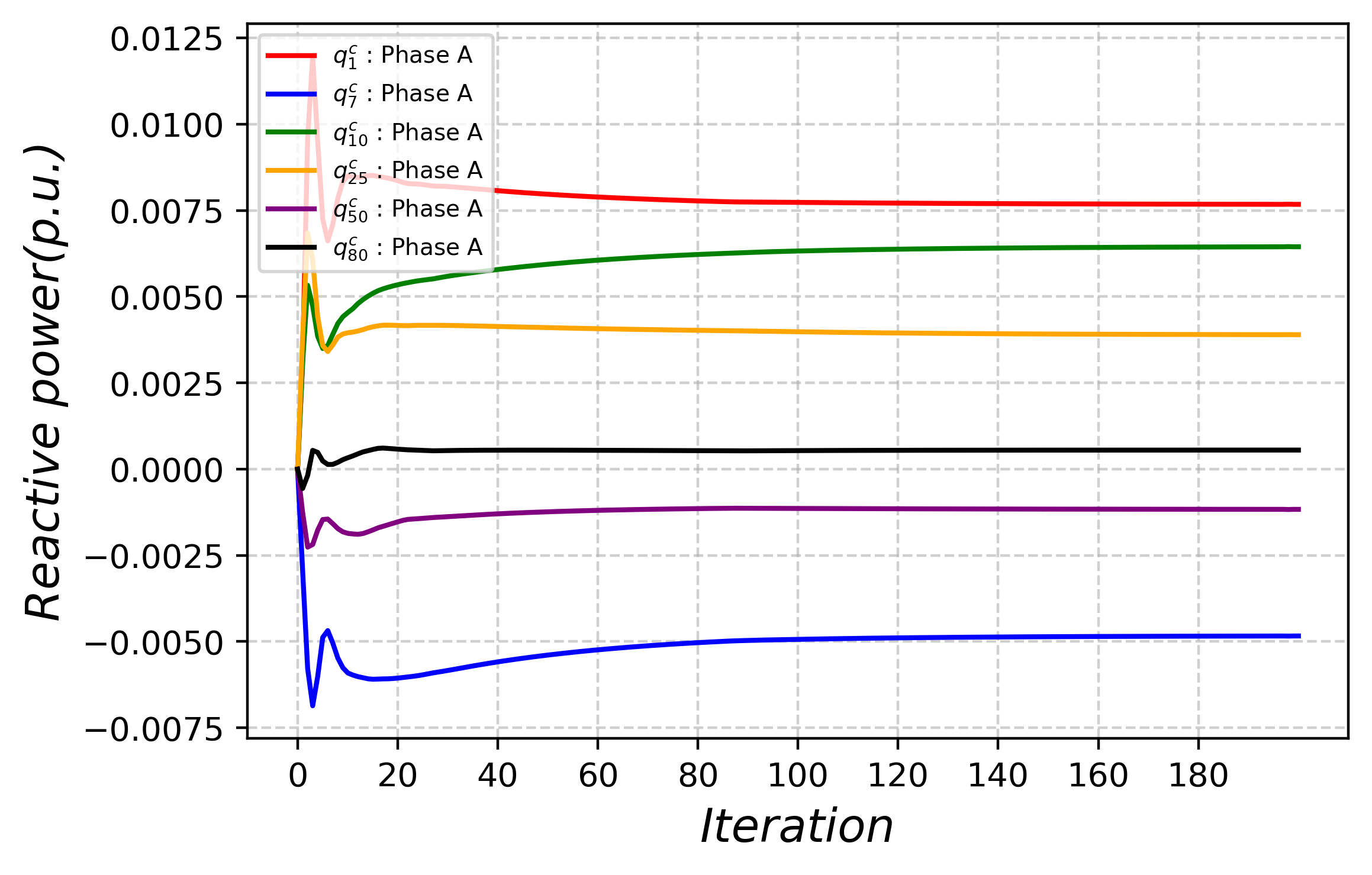}
    \caption{Reactive power outputs versus iterations for various buses 
    by means of ICNN-Assisted OPF in the unbalanced IEEE 123 bus system.}
    \label{fig:ReactivePowerConvergence}
\end{figure}

In the unbalanced three-phase IEEE 123 bus system, we set $C_i^p(\bm{p}_i^c)=\frac{1}{2}\bm{p}_i^c\bm{D}_i^p\bm{p}_i^c$ and $C_i^q(\bm{q}_i^c)=\frac{1}{2}\bm{q}_i^c\bm{D}_i^q\bm{q}_i^c$,  
where $\bm{p}_i^c=[p_{i,\phi}^c]_{\phi\in\Phi_i}$ and  $\bm{q}_i^c=[q_{i,\phi}^c]_{\phi\in\Phi_i}$ denote the real and reactive power vectors of bus $i$, $\Phi_i$ is the phase set of bus $i$, and $\bm{D}_i^p$ and $\bm{D}_i^q$ are diagonal matrices of cost coefficients. And we  utilize the operation data of unbalanced three-phase IEEE 123 bus system to learn (\ref{eq:PQOPF}b)-(\ref{eq:PQOPF}c). The absolute voltage deviation results of unbalanced three-phase IEEE 123 bus system are shown in Fig.\ref{fig:NetworkState}. As shown in the blue icons of   Fig.\ref{fig:NetworkState}, there are voltage violations if we do not take any actions, and the ICNN can also achieve a great prediction for these voltage violations, depicted in the red icons of Fig.\ref{fig:NetworkState}. After utilizing the ICNN-Assisted OPF, there is no any voltage violations in the distribution network, as depicted in the green icons.
Taking buses 1, 7, 10, 25, 50, and 80 as examples, the convergence of phase-A reactive power outputs for these buses are shown in Fig.\ref{fig:ReactivePowerConvergence}. Around 60 iterations, phase-A reactive power outputs for these buses reach convergence. 

The above simulation results show the ICNN-Assisted OPF can ensure 
 the safe and reliable operation of both meshed and unbalanced distribution networks. To further verify the performance of ICNN-Assisted OPF in distribution networks, we consider the following distribution network OPF algorithms:


\textbf{[B1] Supervised Learning for  Distribution Network OPF:} We directly utilize the neural network to learn the mapping from the 
environmental states, e.g., $\bm{p}^{u}$ and $\bm{q}^{u}$, to the optimal decision variables, i.e., $\bm{p}^{c}$ and $\bm{q}^{c}$, by means of supervised learning.

\textbf{[B2] Distribution Network OPF Using Linearized Power Flow:} We utilize the linearized power flow to express (\ref{eq:PQOPF}b)-(\ref{eq:PQOPF}c), then solve it by the commercial solver CPLEX. 

\textbf{[B3] Neural Network-Asssited OPF:} We utilize the neural network model to learn (\ref{eq:PQOPF}b)-(\ref{eq:PQOPF}c), followed by solving (\ref{eq:PQOPF}) based on the fast primal-dual gradient method.

\textbf{[B4] Proposed ICNN-Assisted OPF:} We utilize the ICNN to learn (\ref{eq:PQOPF}b)-(\ref{eq:PQOPF}c), followed by solving (\ref{eq:PQOPF}) based on the fast primal-dual gradient method.

\begin{table}[h]
\normalsize
\centering
\caption{{Distribution Network OPF Comparisons among different algorithms}}\label{tab:OPFCompare}
\footnotesize
\renewcommand\arraystretch{1.0}
        \begin{tabular}{cccccc}
        \hline
        \hline
          Algorithm& $\textbf{B1}$ & $\textbf{B2}$ & $\textbf{B3}$ & $\textbf{B4}$\\
        \hline
        Model-Based&No&Yes&No&No\\
        Safe and Reliable Operation&No&Yes&Yes&Yes\\
        Optimality& No& Yes& No&Yes\\
        Convergence Guarantee& Yes&Yes&No&Yes\\
        \hline
        \hline
        \end{tabular}
\end{table}

\begin{figure}
    \centering
    \includegraphics[width=3.5in]{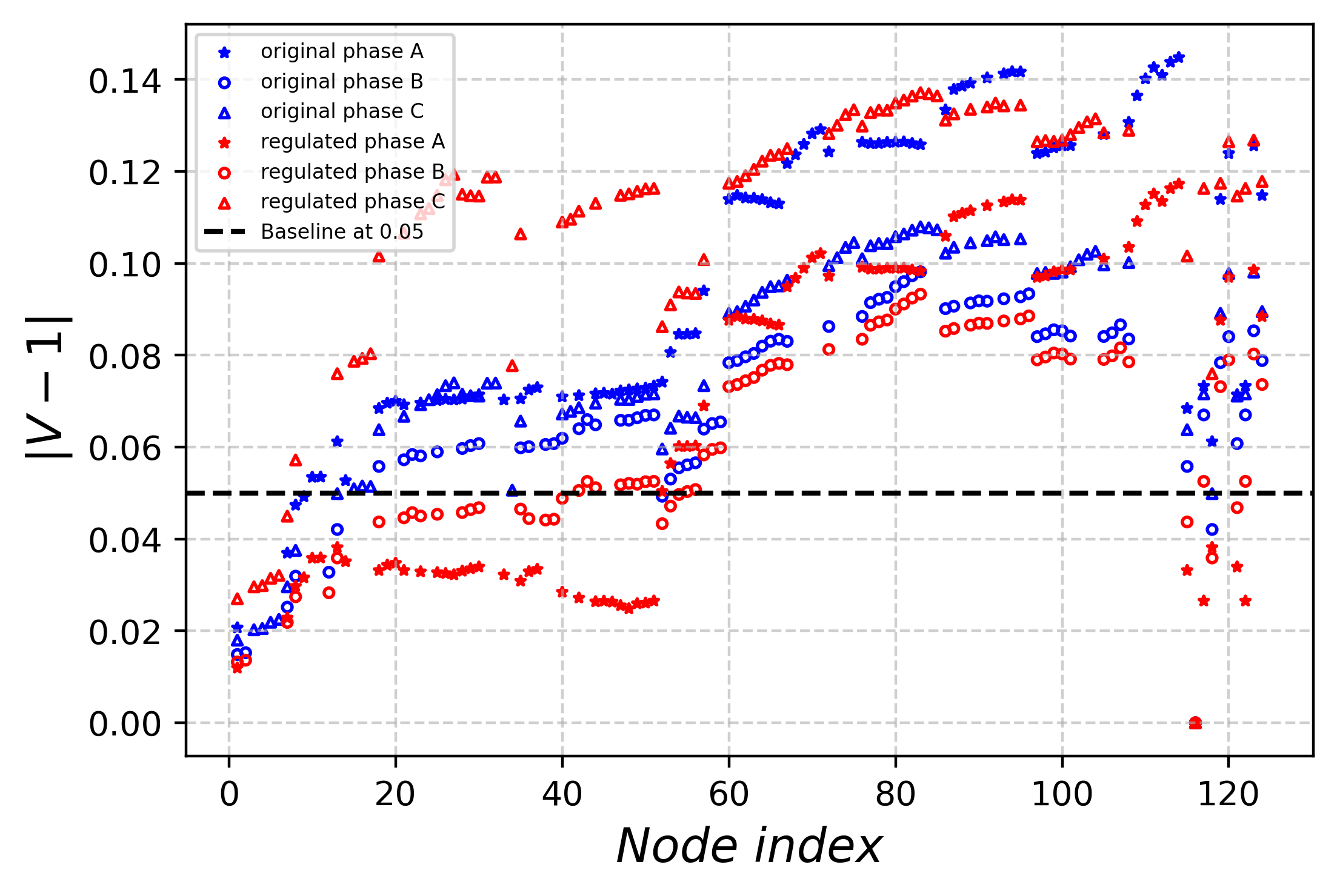}
    \caption{The absolute voltage deviation across the unbalanced IEEE 123 bus system: the blue color denotes the original voltages without any control; the red color denotes the regulated voltage deviations by means of supervised learning}
    \label{fig:SL}
\end{figure}

As shown in Fig.\ref{fig:SL}, \textbf{B1:} supervised learning cannot guarantee the safe and reliable operation of distribution networks. In contrast, other distribution network OPF algorithms are capable of handling distribution network constraints to ensure the safe and reliable operation of distribution networks. \textbf{B2:} Distribution network OPF using linearized power flow are essentially model-based, requiring complete
and accurate distribution network modeling. Unlike \textbf{B2}, \textbf{B3:} Neural Network-Assisted  OPF is data-driven, which does not require physical distribution network models. However, it always suffers from the convergence and optimality problems. In contrast, as depicted in Fig.\ref{fig:ReactivePowerConvergence}. there is no any convergence problem in \textbf{B4:} ICNN-Assisted OPF. This phenomenon is consistent with the theoretical analysis. \textbf{B3:} Neural Network-Assisted  OPF is a non-convex program due to the non-convex properties of neural networks, gradient-based methods cannot guarantee the convergence and optimality performance of such non-convex problems. Instead, \textbf{B4:} ICNN-Assisted  OPF is a convex program. As proved in the previous sections, the  proposed primal-dual gradient method can guarantee the convergence and optimality of ICNN-Assisted OPF in theory.

\section{Conclusion}
\label{sec:conclusion}
This paper proposes the ICNN-Assisted OPF in distribution networks, where its modeling, algorithm design and applications are carefully and comprehensively established, and  We further test its performance on distribution networks of different 
structures, including the meshed and unbalanced distribution networks. The simulation results show that the ICNN can achieve an accurate approximation to the original nonlinear power flow model and exhibit a safe, reliable, and credible performance for distribution network OPF. Future studies will focus on improving the robustness of ICNN-Assisted OPF and extend the ICNN-Assisted OPF to multi-period optimization problems.


\begin{thebibliography}{1}
\bibitem{CMB}
Mary B.~Cain, Richard P.~O’neill, and Anya Castillo. ``History of optimal power flow and formulations.'' \emph{Federal Energy Regulatory Commission}, pp. 1-36, 2012.
\bibitem{FC}
F.~Capitanescu, ``Critical review of recent advances and further developments needed in AC optimal power flow,'' \emph{Electr.~Power~Syst.~Res.}, vol.~136, pp.~57–68, 2016.
\bibitem{PNM}
R.~Cheng, Z.~Wang, Y.~Guo, ``Online voltage control for unbalanced distribution networks using projected Newton method,'' \emph{IEEE Trans.~Power Syst.}, vol. 37, no. 6, pp. 4747-4760, Nov. 2022.
\bibitem{BH}
B. Huang and J. Wang, ``Applications of physics-informed neural networks in power systems-A review,'' \emph{IEEE Trans. Power Syst.},  vol. 38, no. 1, pp. 572-588, Jan. 2023.

\bibitem{ICNN}
B. Amos, L. Xu, J.Z. Kolter, ``Input Convex Neural Networks,'' in \emph{Proceedings of the 34th International Conference on Machine Learning}, 2017, vol. 70, pp. 146-155.

\bibitem{JKA}
J.~Koshal, A.~Nedić, and U. V. Shanbhag, ``Multiuser optimization: Distributed algorithms and error analysis,'' \emph{SIAM Journal on Optimization}, vol~21, no.~3, pp.~1046-1081, 2011.

\bibitem{SG}
S.~Granville, J.~C.~O.~Mello and A.~C.~G.~Melo, ``Application of interior point methods to power flow unsolvability,'' \emph{IEEE Trans.~Power Syst.}, vol.~11, no.~2, pp.~1096-1103, May~1996.

\bibitem{NM}
M.~Niu, C.~Wan, and Z.~Xu. ``A review on applications of heuristic optimization algorithms for optimal power flow in modern power systems,'' \emph{Journal of Modern Power Systems and Clean Energy}, vol. 2, no. 4, pp. 289-297, 2014.

\bibitem{RAJ}
R.A.~Jabr, ``Radial distribution load flow using conic programming,” \emph{IEEE Trans.~Power Syst.}, vol.~21, no.~3, pp.~1458-1459, Aug.~2006.

\bibitem{BFM} 
M.~Farivar and S.H.~Low, ``Branch flow model: relaxations and convexification,'' \emph{IEEE Trans.~Power Syst.}, vol.~28, no.~3, pp.~2554-2564, 2013.

\bibitem{XB}
X.~Bai, H.~Wei, K.~Fujisawa, and Y.~Wang, ``Semidefinite programming for optimal power flow problems,” \emph{Int.~J.~Elect.~Power Energy Syst.}, vol.~30, no.~6-7, pp.~383-392, 2008.

\bibitem{JL}
J.~Lavaei and S.H.~Low, ``Zero duality gap in optimal power flow problem,” \emph{IEEE Trans.~Power Syst.}, vol.~27, no.~1, pp.~92-107, Feb.~2012.

\bibitem{EDAHZ}
E.~Dall'Anese, H.~Zhu, and G.B.~Giannakis, ``Distributed optimal power flow for smart microgrids,” \emph{IEEE Trans.~Smart Grid}, vol.~4, no.~3, pp.~1464–1475, Sep.~2013.

\bibitem{MAS}
M.S.~Andersen, A.~Hansson, and L.~Vandenberghe, ``Reduced-complexity semidefinite relaxations of optimal power flow problems,” \emph{IEEE Trans.~Power Syst.}, vol.~29, no.~4, pp.~1855-1863, Jul.~2014.


\bibitem{MB} 
M.E.~Baran and F.F.~Wu, ``Network reconfiguration in distribution system for loss reduction and load balancing,” \emph{IEEE Trans.~Power Del.}, vol.~4, no.~2, pp.~1401-1407, Apr.~1989.

\bibitem{LG}
L.~Gan and S.H.~Low, ``Convex relaxations and linear approximation for optimal power flow in multiphase radial network,” in \emph{18th Power Systems Computation Conference (PSCC)}, 2014.

\bibitem{RCZ}
R.~Cheng, Z.~Wang and Y.~Guo, ``An online feedback-based linearized power flow model for unbalanced distribution networks,'' \emph{IEEE Trans.~Power Syst.}, vol.~37, no.~5, pp.~3552-3565, Sept.~2022.

\bibitem{KT}
K.~Turitsyn, P.~Sulc, S.~Backhaus and M.~Chertkov, ``Options for control of reactive power by distributed photovoltaic generators," \emph{Proc.~IEEE}, vol.~99, no.~6, pp.~1063-1073, June 2011.


\bibitem{DeepAL}
X. Pan, W. Huang, M. Chen, and S. H. Low, ``DeepOPF-AL: augmented learning for solving AC-OPF problems with a multi-valued load-solution mapping,''  in \emph{ACM e-Energy}, 2023, pp.42-47.

\bibitem{DeepOPF}
W.~Huang, X.~Pan, M.~Chen and S.~H.~Low, ``DeepOPF-V: solving AC-OPF problems efficiently,'' \emph{IEEE Trans.~Power Syst.}, vol. 37, no. 1, pp. 800-803, Jan. 2022.

\bibitem{DDOPF}
X.~Lei, Z.~Yang, J.~Yu, J.~Zhao, Q.~Gao and H.~Yu, ``Data-driven optimal power flow: A physics-informed machine learning approach,'' \emph{IEEE Trans.~Power Syst.}, vol.~36, no.~1, pp.~346-354, Jan.~2021.

\bibitem{UL}
W.~Huang and M.~Chen, ``DeepOPF-NGT: A fast unsupervised learning approach for solving AC-OPF problems without ground truth,'' in \emph{ICML 2021 Workshop on Tackling Climate Change with Machine Learning}, 2021.
\bibitem{DRL}
Y. Zhang, X. Wang, J. Wang, and Y. Zhang, ``Deep reinforcement learning based volt-var optimization in smart  distribution systems,'' \emph{IEEE Trans.~Smart Grid}, vol.~12, no.~1, pp.361-371, 2021.

\bibitem{QZK}
Q.~Zhang, K.~Dehghanpour, Z.~Wang, and Q.~Huang, ``A learning-based power management method for networked microgrids under incomplete information,'' \emph{IEEE Trans.~Smart Grid}, vol.~11, no.~2, pp. 1193-1204.

\bibitem{QZK2}
Q.~Zhang, K.~Dehghanpour, Z.~Wang, F.~Qiu, and D.~Zhao, ``Multi-agent safe policy learning for power management of networked microgrids,” \emph{IEEE Trans. Smart Grid}, vol.~12, no.~2, pp.~1048–1062, Mar.~2021.

\bibitem{WW}
W.~Wang, N.~Yu, Y.~Gao, and J.~Shi, ``Safe off-policy deep reinforcement learning algorithm for Volt-VAR control in power distribution systems,'' \emph{IEEE Trans. Smart Grid}, vol.~11, no.~4, pp.~3008–3018, Jul.~2020.


\bibitem{AVG}
A.~Venzke, G.~Qu, S.~Low, and S.~Chatzivasileiadis, ``Learning optimal
power flow: Worst-case guarantees for neural networks,'' in \emph{Proc.~IEEE
Int. Conf. Commun., Control, Comput. Technol. Smart Grids}, pp.~1–7, 2020.

\bibitem{IM}
I.~Murzakhanov, Venzke, A.~Misyris, G.~S., and S. Chatzivasileiadis, ``Neural networks for encoding dynamic security-constrained optimal power flow,'' in \emph{Proceedings of 11th Bulk Power Systems Dynamics and Control Sympositum 2022}, 2022.

\bibitem{TCL}
G.~Chen, H.~Zhang, H.~Hui, N.~Dai and Y.~Song, ``Scheduling thermostatically controlled loads to provide regulation capacity based on a learning-based optimal power flow model,'' \emph{IEEE Transactions on Sustainable Energy}, vol.~12, no.~4, pp.~2459-2470, Oct.~2021.

\bibitem{ICNNVolt}
Y. Chen, Y. Shi and B. Zhang, ``Data-driven optimal voltage regulation using input convex neural networks,'' \emph{Electr.~Power~Syst.~Res.}, vol. 189, 2020.

\bibitem{DRVVC}
Q. Ma, C. Deng, ``Deterministic and robust volt-var control methods of power system based on convex deep learning,'' \emph{Journal of Modern Power Systems and Clean Energy}, vol.~12, no.~3,  vol. 719~729, 2024.

\bibitem{YZC}
Y.~Chen, Y.~Shi, and B.~Zhang. ``Optimal Control Via Neural Networks: A Convex Approach.'' in \emph{International Conference on Learning Representations}, Sep.~2018.

\bibitem{ASG}
A.~Simonetto and G.~Leus, ``Double smoothing for time-varying distributed multiuser optimization,'' in \emph{2014 IEEE Global Conference on Signal and Information Processing (GlobalSIP)}, Dec. 2014.


\bibitem{testfeeder} 
W.H.~Kersting, ``Radial distribution test feeders,'' in \emph{Proc. IEEE Power Eng.~Soc.~Win.~Meeting}, 2001, pp.~908-912.


\bibitem{MATPOWER}
R.D.~Zimmerman, C.E.Murillo-Sánchez and R.J.Thomas, ``MATPOWER: Steady-State operations, planning, and analysis tools for power systems research and education," \emph{IEEE Trans. Power Syst.}, vol.~26, no.~1, pp.~12-19, Feb.~2011.

\bibitem{OpenDSS} R.~C.~Dugan and T.~E.~McDermott, ``An open source platform for collaborating on smart grid research,” in \emph{Proc. IEEE Power Energy Soc. Gen. Meeting}, Jul.~2011, pp.~1-7.



\end{thebibliography}
\end{document}